\documentclass[manuscript]{acmart}

\AtBeginDocument{%
  \providecommand\BibTeX{{%
    \normalfont B\kern-0.5em{\scshape i\kern-0.25em b}\kern-0.8em\TeX}}}

\usepackage{xcolor,colortbl}
\definecolor{light-gray}{gray}{0.95}
\newcolumntype{a}{>{\columncolor{Gray}}c}
\usepackage{url}
\usepackage[normalem]{ulem}
\usepackage{subcaption}
\usepackage{tablefootnote}

\begin{document}

\title{Runtime Software Patching: Taxonomy, Survey and Future Directions}
\author{Chadni Islam}
\authornote{The first and second authors contributed equally to this research.}

\email{chadni.islam@adelaide.edu.au}

\author{Victor Prokhorenko}
\email{victor.prokhorenko@adelaide.edu.au}

\author{M. Ali Babar}
\email{ali.babar@adelaide.edu.au}
\affiliation{%
  \institution{University of Adelaide}
   \city{Adelaide}
   \country{Australia}
  \postcode{SA 5005}
 }

\begin{abstract}

Runtime software patching aims to minimize or eliminate service downtime, user interruptions and potential data losses while deploying a patch. Due to modern software systems' high variance and heterogeneity, no universal solutions are available or proposed to deploy and execute patches at runtime. Existing runtime software patching solutions focus on specific cases, scenarios, programming languages and operating systems. This paper aims to identify, investigate and synthesize state-of-the-art runtime software patching approaches and gives an overview of currently unsolved challenges. It further provides insights into multiple aspects of runtime patching approaches such as patch scales, general strategies and responsibilities. This study identifies seven levels of granularity, two key strategies providing a conceptual model of three responsible entities and four capabilities of runtime patching solutions. Through the analysis of the existing literature, this research also reveals open issues hindering more comprehensive adoption of runtime patching in practice. Finally, it proposes several crucial future directions that require further attention from both researchers and practitioners.

\end{abstract}

\maketitle

\section{Introduction}\label{sec:Intro}

Modern software is constantly evolving and adapting to ever-changing user needs and requirements, leading to the necessity to apply various changes to the existing software code. In the traditional software development life cycle, software changes are implemented in code compilation and deployment steps. These changes range from simple bug fixes to full-fledged software reworks. The increasing complexity of modern software and growing user demands lead to frequent software modifications that are provided in the form of updates and patches. Such code modifications typically attempt to improve or extend the existing software functionality to satisfy new user requirements. Improving does not necessarily relate to the core software system functionality; however, it can focus on enhancing the auxiliary properties such as security or privacy. 

There is no universally accepted formal definition of the difference between patching and updating. However, a common understanding is that software changes that introduce new functionality are called \textit{updates} (or upgrades). In contrast, minor changes that fix existing bugs or vulnerabilities are referred to as \textit{patches}. Alternatively, some versioning systems\footnote{\url{https://semver.org/}} loosely define software compatibility as a dividing line between patches and updates. However, even minor bug-fixing patches may break the compatibility with the previous version, rendering such compatibility-based distinction inadequate in practice. Regardless of the terminology used, both patches and updates essentially refer to \textbf{code modifications}, which range from \textit{simple} to quite \textit{complex}.

In the simplest form, replacing the previously running software instance with an updated version involves stopping the old software instance and starting the new one~ \cite{inst8-DSU-Hics2005, Inst1-Karmachen2017adaptive, Function7Kshot, Function1KspliceArnold2009, Process8-hayden2012kitsune}. Such running software instances could be individual processes, web services, virtual machines or containers. However, depending on the extent of the change and type of the software, the update process may cause lengthy software service downtime, which can negatively impact end-users' experience. From an end-user perspective, updates can also be perceived as either existing usage session interruptions or new session establishing delays. Such software service disruptions may cause severe negative consequences in highly-critical environments such as health or industrial control domains. In addition, highly loaded profit-oriented environments such as large data centers or stock exchanges may suffer direct financial losses from even short downtime\footnote{\url{https://www.upguard.com/blog/the-cost-of-downtime-at-the-worlds-biggest-online-retailer}}. A recent report by Gartner has indicated the cost of downtime could be~\$1-5 million at the higher end for just one hour~\cite{Gartner}. Therefore, the notion of \textit{runtime software patching} (in short runtime patching) is developed, which is also referred to as hot-patching, dynamic patching, or live patching~\cite{Function7Kshot}. 

In an attempt to improve traditional (also known as offline) patching, runtime software patching aims to minimize or completely avoid any software operation interruptions. Runtime patching is commonly considered a technique for fast vulnerability mitigation~\cite{inst8-DSU-Hics2005, Function1KspliceArnold2009, Inst5-chen2013safestack, Inst4-chen2018instaguard}. However, compared to simple traditional patching (i.e., source code patching), patching a running software poses significant challenges. Namely, preserving existing end-user activities (as well as compatibility with future ones) ranging from human input to complicated network-based communication sessions must be achieved for efficient runtime patching. In other words, the future behavior of the software must be affected by the applied patch, with currently executed activities and used data also needing to be adjusted accordingly to maintain compatibility. In some cases, runtime patching injects code into vulnerable programs to achieve a temporary fix~\cite{Inst5-chen2013safestack, Inst4-chen2018instaguard}. These temporary fixes are designed to immediately disable or replace vulnerable code to prevent the exploitation of the vulnerability. In the meantime, a developer can implement and test a proper update that actually fixes the bug, rather than just disabling the currently broken functionality. Note that additional non-technical supply chain requirements related to patch approval, testing, signing and distribution may introduce further significant patch adoption delays in practice~\cite{Inst4-chen2018instaguard}.

Adopting runtime patching requires intimate knowledge of the software implementation details. For instance, converting existing internal functions or objects to be compatible with the updated software version requires a deep understanding of the object formats, structures and locations. However, no comprehensive runtime patching mechanisms have been implemented due to the technical diversity of underlying hardware, compilers, third-party libraries, and Operating Systems (OS). Existing runtime patching solutions target specific domains or scenarios such as IoT devices or virtual machines and hypervisors at datacenter scale~\cite{Inst6-Function5-Replus-chen2015framework, Function10-salls2017piston, Process1-mugarza2020cetratus, hypervisor2-Orthus-zhang2019fast, VM2-zhang2014vpatcher}. These solutions typically focus on platform-specific challenges such as hardware limitations inherent to resource-constrained IoT devices. For instance, even storing the patched copy of software might not be possible under tight storage constraints.  


Existing patching studies and reviews do not provide a formal approach to determining patch scale quantitatively, hence, leaving a gap in terminology related to patch and update differences. In addition, due to the high number and diversity of existing platform-specific patching solutions, a lack of comprehensive and formal understanding of general runtime patching strategies and required implementation capabilities such as approach applicability and expected potential disruptions are noticed. Moreover, the roles of different parties involved in runtime patching and their responsibilities in various patching process steps are also not explicitly considered.


Therefore, we review the state-of-the-art literature on software runtime patching from the deployment aspects to gain an understanding of existing runtime patching techniques and approaches. Based on the analysis, we propose a taxonomy (discussed in Section~\ref{sec:runtimePatching}) that provides a comprehensive view of the runtime patching solutions from the perspective of patch granularity (Section~\ref{sec:Granularity}), patch approaches (Section~\ref{sec:patchStrategiesCapability}) and patch responsibilities (Section~\ref{sec:PatchResponsibility}). 
We also identify the challenges and the research gaps in runtime patching and highlight the future research direction for further improvement in this field in Section~\ref{sec:futuredirection}. 

    

    
    


The rest of this study is structured as follows. Section~\ref{sec:Background} discusses the software patching in general, followed by section~\ref{sec:runtimePatching} diving further into an overview of runtime patching specifically. Various runtime patching granularities are examined in Section \ref{sec:Granularity}. A comprehensive view of existing runtime patch deployment strategies, workflow and capabilities is presented in Section \ref{sec:patchStrategiesCapability}. Section \ref{sec:PatchResponsibility} contains a detailed discussion of parties involved in the patching process along with their corresponding roles and responsibilities. Further research opportunities addressing the open challenges identified are outlined in Section \ref{sec:futuredirection} with Section \ref{sec:conclude} concluding this study.

\section{Background}\label{sec:Background}

\begin{figure*}[htb]
  \centering
  \includegraphics[scale=0.8]{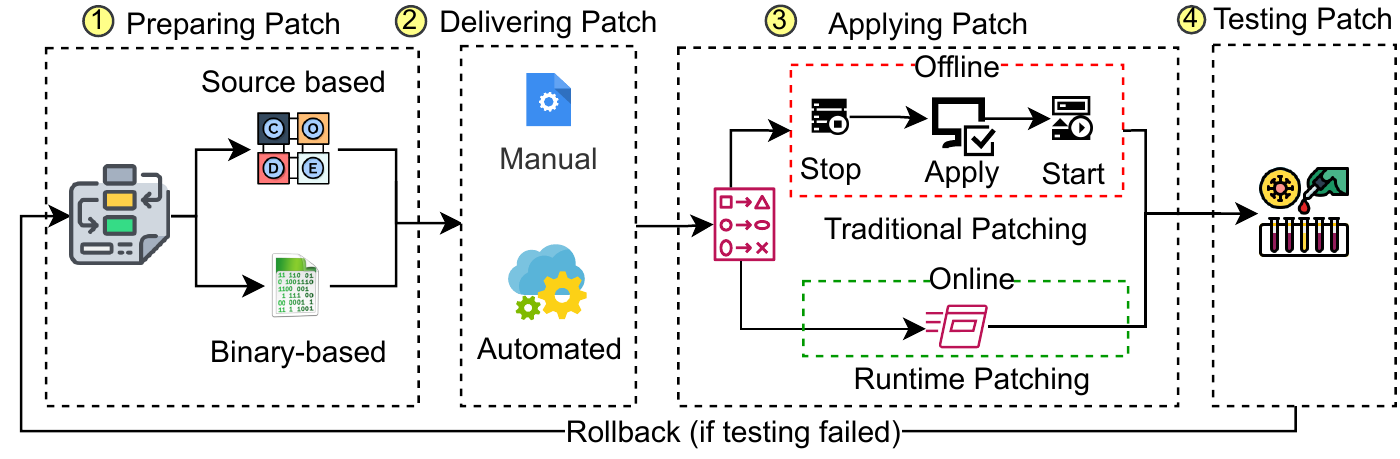}
  \caption{The key steps of software patching life cycle.}
  \label{fig:life_cycle}
\end{figure*}

A generic patch management process is shown in Figure~\ref{fig:life_cycle}. Irrespective of the extent of a change, updating a software system can be performed offline or online. As can be seen from Figure~\ref{fig:life_cycle}, patch preparation, delivering a patch, applying a patch, testing the patch and potentially removing the patch (if applicable) are the key steps. Techniques of patch preparation differ depending on whether the original source code is available. Source-code patches are essentially self-contained and logically united code edits to improve or fix certain functionality, bugs or vulnerabilities. Modern code version control systems typically track such edits as code commits. Having full access to application source code enables the development of comprehensive and flexible changes. However, the main downside of source-based patching is the requirement to recompile the whole application to include the required changes. Depending on the size of the software package, the recompilation step alone might take significant time.

In contrast, lack of source code access means that the binary executable must be altered directly. Therefore, binary-based patches essentially contain the bytes that need to be changed in the original binary executable. Compared to source code patching, binary patching generally takes a shorter time as no lengthy recompilation step is necessary. However, binary patching poses an additional challenge during the preparation steps. Namely, the addresses of the bytes to be altered must be located within the original binary.

While outside of the main scope of this review, it is worth noting that rollback planning is crucial for all changes occurring in a production system. It is typically assumed that rollback occurs when the new behavior does not match the expected changes. This assumption may mean that some downtime is inevitable. Thus, rollback procedures can be considered somewhat less constrained in their nature. When possible, however, maintaining operation is still a priority while the rollback is being conducted. Therefore, in general, a rollback operation is equivalent to a runtime patch in the opposite direction (new code back to old code). The same self-consistency considerations must be taken into account in both directions. The key difference is that a longer preparation is possible before a patch is applied compared to potential abrupt failures. Generally, a rollback procedure highly depends on the type of patch applied, application data involved, expected impact, timing constraints and amount of code. This review purports to focus only on the patch deployment steps, approaches and solutions.

\vspace{-5pt}

\subsection{Traditional Software Patching}

Traditional or offline patching implies stopping the running software instance, potentially converting existing data and starting a new instance~\cite{Inst6-Function5-Replus-chen2015framework, Function10-Orso2002DUSC}. As shown in Figure~\ref{fig:life_cycle}, the first step in software patching is patch preparation. The prepared patch targets \textbf{source} or \textbf{binary levels} depending on whether the source code is available. Source-targeted patches require a potentially time-consuming software recompilation to produce a new (patched) executable, while binary-targeted modify the existing software executable. Source code level patching allows the programmer to easily modify arbitrary aspects of the software, such as replacing any functions, instructions and data flows. On the other hand, binary patches are somewhat more limited in their nature and are used for smaller-scale modifications (primarily in the security context). The main complication in binary patching is that any code length changes can cause significant memory pointers invalidation. In addition, in binary-level patching, finding the exact location to modify the buggy part may be complicated~\cite{xu2020automatic-Kernel4F-Inst2Vulmet}. Besides, the old binary can still run, further complicating the patching process. Furthermore, as the newly patched binary is ready for execution, running it may cause conflict due to resource sharing when the old binary uses the same resource. 

\textbf{Challenges with traditional patching}:
The major drawback of the traditional software patching approach is the associated \textbf{service interruption} when the old binary is stopped and the new one is not fully started \cite{inst8-DSU-Hics2005, Function10-Orso2002DUSC}. For instance, remote or local user sessions, network connections and data processing are interrupted or suspended when the patched software system stops. Short interruptions may be acceptable (albeit inconvenient) in interactive user sessions such as web browsing or word processing. Interactive software is commonly designed to automatically save and restore user sessions to mitigate such inconveniences to some extent. Web browsers that attempt to restore the state between restarts are a good example of such mitigation.
In contrast, experiencing \textbf{service disruptions} in highly critical systems may cause significant monetary losses or be considered entirely unacceptable. For example, for life-support software and air-traffic controller, shutting down a system is prohibited or not considered an option~\cite{Function10-Orso2002DUSC}. Thus, system administrators may opt to delay patch deployment to avoid potential system reboots that cause service disruption and loss of application state. 
For example, updating the OS on servers for highly interactive activities like online gaming or video streaming typically needs to schedule server downtime. During this time, players have to stop the game, wait for the servers to be updated and restarted, then login back in to the server and potentially start the game from the beginning, which is bothersome for the players. Unfortunately, keeping a vulnerable system un-patched for a prolonged period of time increases the risk of having the system exploited. 
In efforts to overcome these challenges of traditional patching, runtime patching has emerged and gained popularity. Details of runtime patching and related benefits are discussed in Section~\ref{sec:runtimePatching}.

\subsection{Existing reviews}\label{sec:ExistingReview}

Several software patch and update related reviews have been conducted in recent years~\cite{seifzadeh2013survey, ahmed2020dynamic-LR, miedes2012survey, lopez2017survey, ilvonen2016dynamic, gregersen2013state, mugarza2018analysis}. Ahmed et al.~\cite{ahmed2020dynamic-LR} have performed a comprehensive systematic mapping study related to runtime software updating solutions. We encountered somewhat unstructured approach classifications while analyzing a large number of papers. For example, the most cited updating approaches include ``Java VM" and ``Multi-version," which do not necessarily have to be mutually exclusive. Furthermore, while the study has provided the statistics and an overview of approaches used, it lacks details on approach-specific benefits and challenges and the correlation between the adoption of techniques, tools and algorithms.

A more structured categorization of runtime software updating solutions is presented by Seifzadeh et al.~\cite{seifzadeh2013survey}. This study presents a comprehensive, albeit high-level set of runtime updating evaluation metrics, such as scope, time of update, and type safety. Rather than focusing on existing implementations, Miedes et al.~\cite{miedes2012survey} have outlined the concepts and techniques used in runtime software updating in general. In addition, the set of goals and requirements such as service continuity and generality are identified and discussed. However, no coherent taxonomy or classification of such approaches is presented, leading to somewhat mismatched categories. For instance, while being orthogonal in terms of goals, Java-oriented approaches and rollback-ability are discussed alongside technical challenges. 

In addition, some reviews focus on particular technical domains and usage scenarios. For instance, Lopez et al.~\cite{lopez2017survey} have surveyed existing function and system call hooking approaches. While function hooking can be used for a multitude of purposes (including malicious), function-level patching can significantly benefit by applying hooking techniques. One of the main strengths of the survey is the comprehensive view of both function and system call hooking under major operating systems. 
However, the scope of this review is limited to function granularity only and does not consider other patching levels. Another study by Gregersen et al.~\cite{gregersen2013state} have compared three existing runtime patching implementations for Java applications. On top of evaluating the performance of the implementations, low-level patching capabilities were analyzed. The scope of the review is only limited to Java-specific capabilities such as class modifications were considered. Lastly, Mugarza et al.~\cite{mugarza2018analysis} have focused on runtime software updating in the industrial IoT domain. Specifically, the requirements of safety systems are evaluated using nuclear control systems as a case study. 

Unlike the existing reviews, Ilvonen et al. ~\cite{ilvonen2016dynamic} have taken an interesting perspective by analyzing the support of runtime or Dynamic Software Updating (DSU) in the software engineering education context. In particular, existing software engineering courses are analyzed to determine the adoption and coverage of DSU concepts. The main finding of this study is the lack of a holistic approach towards DSU in education, with only certain individual aspects being addressed in education. 

Several observations are made based on the reviewed studies. Firstly, there is still a lack of common understanding in the domain of runtime patching, even at the level of terminology. Secondly, no clear taxonomy coherently categorizing the existing runtime patching approaches has been developed so far. Thirdly, lack of generality hinders the wider adoption of the existing runtime software patching methodologies and tools. Fourthly, evaluation metrics vary wildly depending on the intended runtime patching domain, ranging from generic time overhead to language-specific class modifications.

\textbf{Scope of our survey:} Our survey focuses on multiple aspects of runtime patching approaches such as patch scale, general tactics and responsibility. A detailed taxonomy is presented along with the corresponding analysis of the existing solutions. The main focus lies within the \textit{applying patch} phase in the patch management life cycle shown in Figure~\ref{fig:life_cycle}. Unlike the existing surveys, which are typically narrow-focused and mainly highlight the methodologies and tools used for patching, we attempt to generalize the issues and approaches inherent to different patch granularity levels. It is worth noting that our survey does not compare the performance of different techniques or solutions because of the vastly different experimental setups and execution environments observed. Similarly, due to the high diversity of the solutions reviewed, different evaluation techniques cannot be compared directly, as different implementations focus on different metrics (e.g., downtime, overhead and long-term patch continuity). Naturally, systems aiming to achieve zero downtime would not even consider such a metric in their evaluation procedures. Furthermore, we indicate the existing state-of-the-art solutions' common challenges and suggest a set of future directions to advance the field. 

\section{Runtime Patching}\label{sec:runtimePatching}
Runtime software patching aims to update a given software system while preserving running processes and sessions. In case some downtime is inevitable, runtime patching approaches focus on minimizing the disruption time. Zhou et al.~\cite{Function7Kshot} has defined runtime patching as "\textit{a method for dynamically updating software, effectively reducing the downtime and inconvenience often associated with software upgrades}". In contrast to traditional patching, runtime patching is primarily binary-oriented because the running binary instance of a program is modified in memory. The binary representation of the code needs to be replaced at runtime with the new (i.e., patched) version~\cite{Kernel1Kup, Function7Kshot}. In addition to patching the in-memory version of a binary, a disk copy must be patched correspondingly so that patched behavior persists in any future restarts.

Runtime patching must take care of the current state that needs to be transformed to be compatible with the new code~\cite{Function8-Cure-zhao2016, Function6-chen2007polus, pina2016tedsuto} that includes in-RAM objects, data structures and external OS resources. A set of existing approaches address various aspects of running state transformation such as update points and state transformers~\cite{Inst5-chen2013safestack, Inst4-chen2018instaguard, Process3-giuffrida2013safe-Proteos}. In such approaches, the patch consists of a combination of the new code, safe update points and necessary state transformers. Update points are essentially the time windows suitable for applying a runtime patch. Data-specific state transformers can then transform the current program state to a new version. Specifically, a runtime patching system continuously monitors the program execution. When (if) a program reaches a suitable update point, the system loads the patched code and starts the program state transformation according to the specified state transformers. Once the transformation is complete, program execution continues with the new version being active~\cite{Function8-Cure-zhao2016}. 

In severe cases, typically for large and complex systems, the state transformation may take considerable time, inducing noticeable service disruptions. For instance, transforming an existing Virtual Machine (VM) to make it compatible with a new hypervisor version might involve converting the virtual disk format. Given the multi-gigabyte sizes of modern disks, this operation might take a significant amount of time. Furthermore, not stopping the VM during the conversion could cause the disk contents to change before the new disk image is finalized. In such cases, runtime patching may choose to perform the conversion in multiple steps and only stop the VM during the last data portion conversion. 

\subsection{Goals and Benefits of Run-time Patching}

Runtime software patching approaches provide a number of practical benefits. First, from the security perspective, patching a running software reduces the vulnerable time window while a long-term solution is being prepared. A typical example of such immediate fixes is temporarily disabling a vulnerable code path, with the long-term solution to fix the expected code behavior. In some cases, applying such simple patches gain an additional time necessary for long-term testing changes at the expense of reduced functionality. Some prominent goals and  benefits of runtime patching include reducing \textit{service interruption}~\cite{inst8-DSU-Hics2005, Function10-salls2017piston, Inst7-payer2013dynsec}, \textit{system downtime}~\cite{ Function7Kshot, Function10-salls2017piston}, \textit{frequency of reboots}~\cite{Inst7-payer2013hotAsap, Function1KspliceArnold2009, hypervisor2-Orthus-zhang2019fast} and \textit{human involvement}~\cite{Function10-salls2017piston, Function1KspliceArnold2009}. While achieving these goals, a runtime patching solution also aims to avoid loss of data and running  states of applications to improve end-user experience~\cite{ Function7Kshot, Kernel1Kup, hypervisor1-hyperFresh-doddamani2019, Function1KspliceArnold2009}. In combination, these advantages of runtime software patching are beneficial from a business perspective in terms of service reliability and reduction of potential monetary losses. Reviewing the existing study, we identify that \textit{high availability}, \textit{quick vulnerability mitigation} and \textit{improving user experiences} are three key aspects in consideration behind runtime patching. 

\vspace{-5pt}

\begin{itemize}
    \item \textbf{High availability}: Providing uninterrupted highly-available service is crucial in critical domains such as defense, medical, industrial control systems and cloud systems. As such mission-critical systems require high availability~\cite{inst8-DSU-Hics2005, Inst5-chen2013safestack, Inst6-Function5-Replus-chen2015framework, Function8-Cure-zhao2016, hypervisor2-Orthus-zhang2019fast}, any disruptions caused by applying patch are unacceptable.
    \item \textbf{Quick vulnerability mitigation}: Runtime patching allows to mitigate vulnerabilities by fixing or disabling unsafe code fragments on the fly~\cite{Inst5-chen2013safestack,  Inst1-Karmachen2017adaptive, Inst4-chen2018instaguard, Function1KspliceArnold2009}. In addition to full-fledged logic improvements, quick temporary solutions such as disabling vulnerable code paths or filtering unsafe user input can be used in practice.
    \item \textbf{Improving user experience:} Applying proper long-term runtime patches reduces the number of software system restarts, thus effectively simplifying overall system maintenance. In addition, this improves the end-user experience as fewer user workflow disruptions (including potential recovery steps) would be encountered during day-to-day system usage~\cite{inst8-DSU-Hics2005, Inst1-Karmachen2017adaptive, xu2020automatic-Kernel4F-Inst2Vulmet}. From an end-user perspective, runtime patching attempts to reduce either frequency or length of service disruptions.
\end{itemize}

 


\subsection{Proposed Taxonomy}

\begin{figure}[tbh]
  \centering
  \includegraphics[scale=0.92]{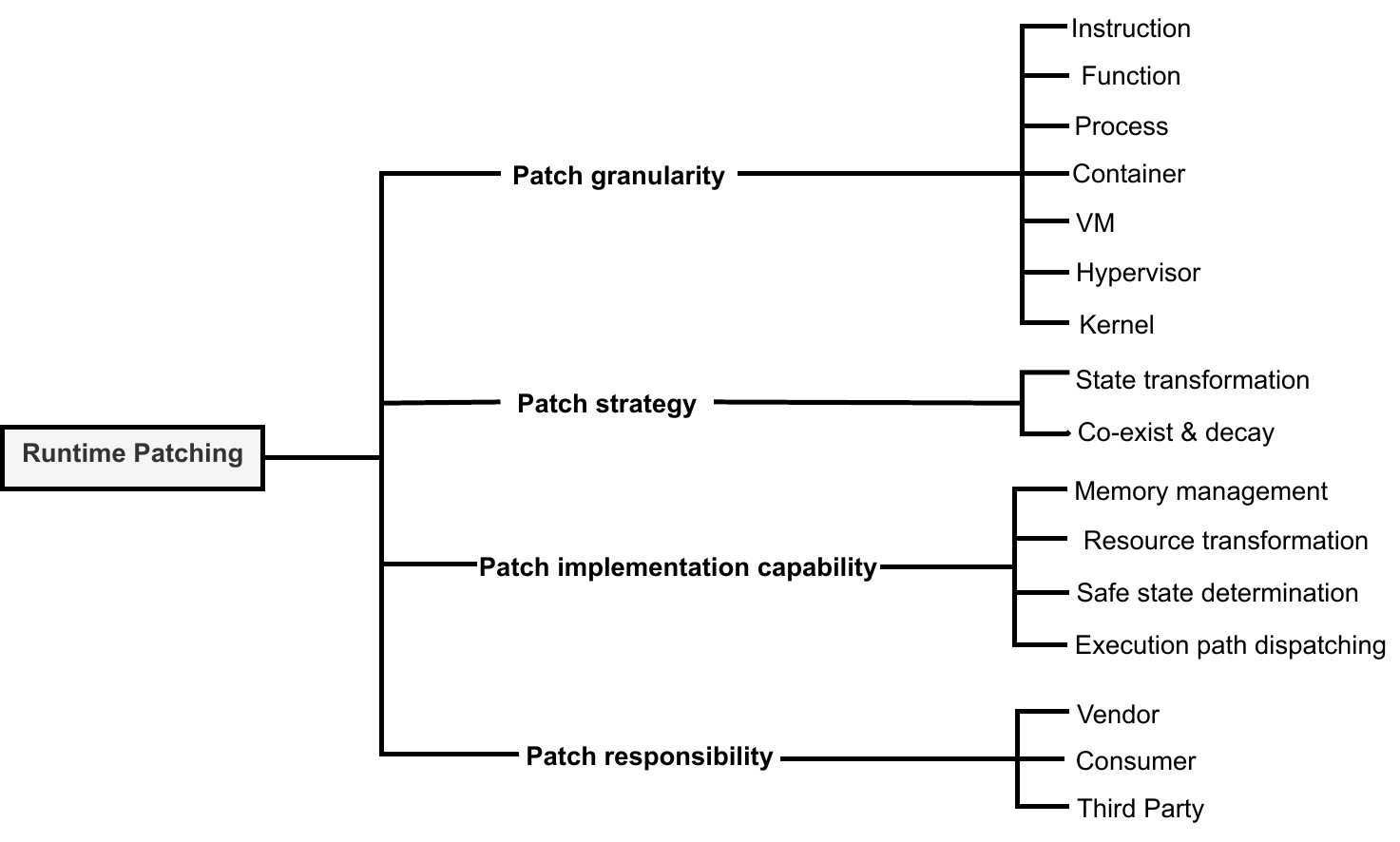}
  \caption{Taxonomy of runtime patching}
  \label{fig:Patchtaxonomy}
  \vspace{-5pt}
\end{figure}

We propose a taxonomy to categorize the studies related to runtime patching as shown in Figure \ref{fig:Patchtaxonomy} in terms of \textit{what}, \textit{how} and \textit{who}. While analyzing the patching techniques incorporated in different studies, we identify that patches are applied at different levels, such as individual function, single process or the whole Operating System~(OS) kernel. We further find that a number of patching levels are implemented in the existing approaches. Hence, the first category we have in our taxonomy is \textbf{patch granularity}. By patch granularity, we mean the scope of a patch supported by a given solution. The scope of a patch refers to the scale of the change supported by a patching solution. For instance, some solutions support patching individual machine-level \textit{instructions}, while others support replacing whole \textit{functions}. On the other hand, larger-scale approaches focus on updating higher-level units such as whole \textit{containers} or \textit{VMs}. 

We notice that when considering what to patch, the surveyed studies have addressed different issues and proposed different approaches. Therefore, analyzing the studies, we extract seven granularity levels by grouping the related runtime patching studies in terms of granularity that the studies aim to patch at runtime. Note that only a few papers are categorized into more than one sub-category for cases where one study has proposed an approach that performs patching at multiple granularities~\cite{Inst6-Function5-Replus-chen2015framework}.

The second and third categories focus on the "how" aspect of the patching and the general techniques employed by an approach (\textbf{patch strategy}) and technical capabilities of a given patching system implementation (discussed in Section~\ref{sec:patchStrategiesCapability}. Two main strategies identified are \textit{state transformation} and \textit{co-exist~\&~decay}. In some cases, direct state transformation is possible when the changes imposed by a patch are not significant. For instance, adding an object property or a method to reflect new (patched) functionality could be straightforward. In contrast, removing or modifying existing functions is likely to cause other code fragments that rely on previous behavior. Thus, the co-exist \& decay approach aims to separate old and new data objects based on sessions or transactions where possible. In other words, old (unpatched) objects currently in use are not modified, while new code patches would be directed at updated objects. Later, old objects are disposed of when the pre-existing sessions/transactions are completed. The \textbf{implementation capabilities} directly reflect the practical applicability of a given solution. For instance, some systems might be capable of automated safe updating time windows detection, while others lacking such capability have to resort to manual assistance from the developers. Similarly, less capable systems may require extra external assistance in memory management to load or fit patched code into RAM.

In the fourth category, we have identified the entities involved and their \textbf{responsibilities }considered by a patching system. Specifically, we focus on "who" is responsible for applying the patches. The existing studies take three common responsibility-targeted views. First, \textit{vendor-supported} patching systems that imply the original software system developers as in charge of the patching process. Second, \textit{software system end-users} who need to patch a running system. Lastly, independent \textit{third-party patchers} who attempt to provide facilities to apply generic patches to existing generic software (within certain limits). These situations differ in the amount of prior knowledge available to the patchers. Most significantly, developers would naturally access the original software source code, while end-users would not. Similarly, original developers would possess more profound knowledge of the application internals and logic.

There can be other ways the studies can be categorized; however, our primary focus is to identify the level at which a patch is applied and the adopted or proposed approaches to apply the patch at runtime. This helps us determine the issues of runtime patching approaches being neglected by the practitioners and identify potentially beneficial future directions.

\vspace{-5pt}

\section{Runtime Patching Granularities}\label{sec:Granularity}

\begin{figure}
    \centering
    \includegraphics[scale=0.7]{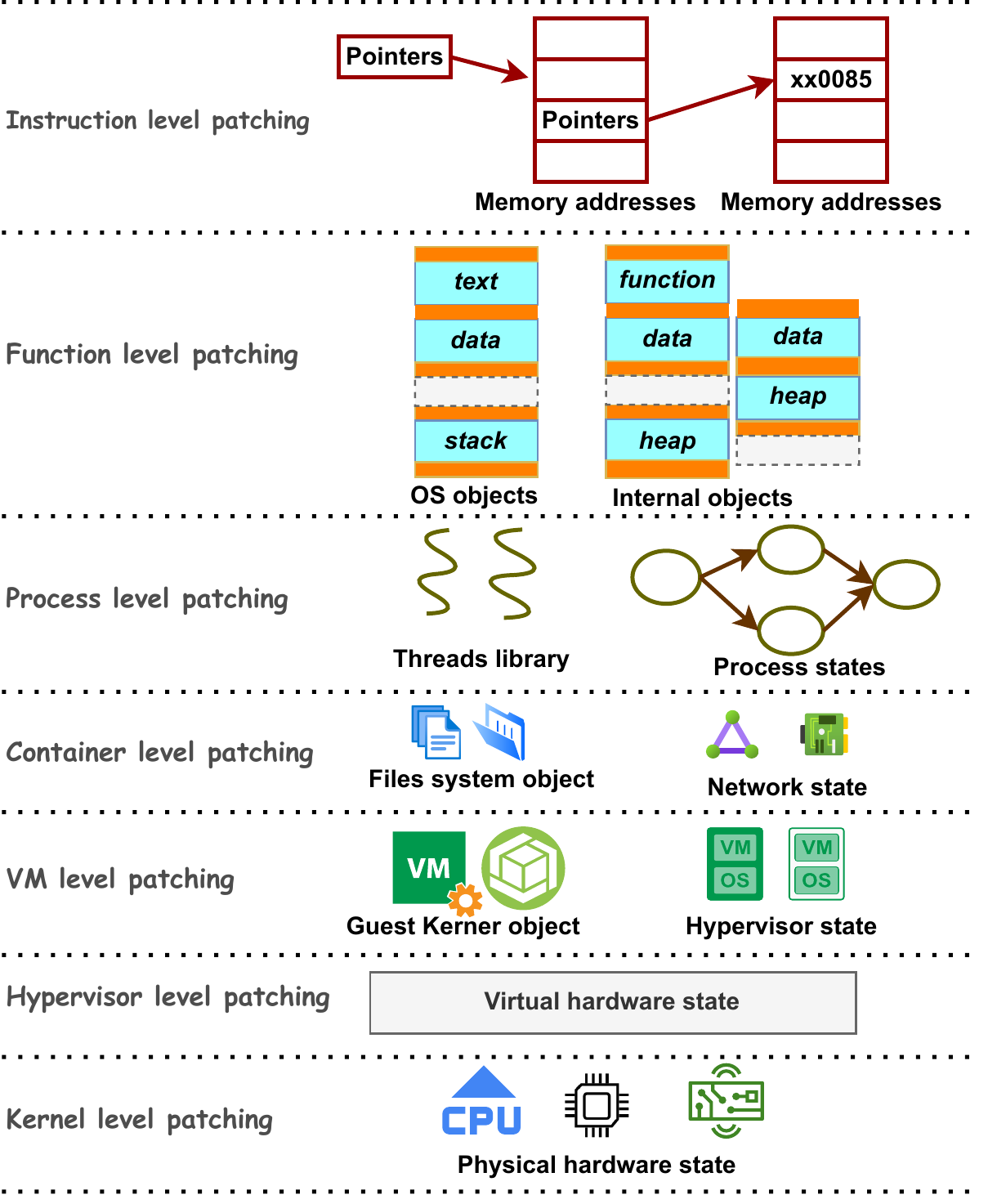}
    \caption{Granularity-specific runtime assets, objects and patch boundaries. Patching at a lower layer of granularity requires transforming or dispatching the assets and objects on the same and all above layers.}
    \label{fig:patchlevel}
\end{figure}

This section covers the intended \textit{patch granularity} in runtime patching that essentially characterizes the scale and boundaries of a patch to be applied. Patch granularity defines a responsibility boundary within which no state transformations are typically required or performed. The studies we have identified are categorized into seven groups: Instruction-level, Function-level, Process-level, Container-level, VM-level, Hypervisor-level and Kernel-level. Figure~\ref{fig:patchlevel} provides an overview of the objects or elements involved in patching at different granularities. While most straightforward patches target changing individual machine instructions in RAM, complex patches target replacing a vulnerable VM with a fixed one. For instance, Figure~\ref{fig:patchlevel} shows that patching an instruction requires taking memory pointers while patching a Host OS requires modifying the physical hardware state into account. Categorizing studies in terms of patch granularity gives insight into the objects and elements that require modifications when applying patches. It further helps understand the impact of a patch (i.e., the extent of change/patch consequences). For instance, changing an object in a lower layer, e.g., in a physical hardware state, impacts the higher layer object, such as file system and OS objects.

The patch preparation and deployment process highly depends on the patch granularity. For instance, replacing a few CPU instructions involves merely compiling the corresponding code into a target CPU architecture, whereas updating a VM may require rebuilding the whole VM image. Technical issues related to each granularity are discussed below. Unless significantly affecting the overall function behavior, patching individual instructions only takes care of the instruction abstraction level. Similarly, replacing a function with a patched variant needs to focus on the function abstraction level unless the overall process behavior change is externally observable.
In contrast, switching over to a new kernel would need to take care of all processes that are using old kernel resources. Essentially, choosing a suitable patch granularity is a trade-off between transforming internal and external states. As shown in Figure~\ref{fig:patchlevel}, connectivity points outside of the set boundary such as network sockets, external function calls and hardware may need to be adjusted to match the patched code.

\subsection{Instruction-level Patching}

Instruction-level patching requires the patching solution to write the changed instructions into an appropriate memory location. It aims to replace machine-level CPU commands or individual bytes in the RAM~\cite{Inst1-Karmachen2017adaptive, xu2020automatic-Kernel4F-Inst2Vulmet, Inst3-WordPatch-chamith2016living}. We have identified a set of studies that have performed instruction level patching~\cite{Inst1-Karmachen2017adaptive, xu2020automatic-Kernel4F-Inst2Vulmet, Inst3-WordPatch-chamith2016living, Inst4-chen2018instaguard, Inst5-chen2013safestack, Inst6-Function5-Replus-chen2015framework}. Patch preparation slightly differs in instruction level patching depending on the source code availability for a given software. In both cases, the patch preparation output is a sequence of bytes to be written to a certain memory address (potentially computed dynamically). Instruction level patching is lightweight and faster than function level patching~\cite{Inst6-Function5-Replus-chen2015framework}.

Table~\ref{tab:instPatching} summarizes the works that have proposed instruction-oriented runtime patching approaches. We have identified the goal of each study along with the task related to patchings, such as patch generation, deployment, verification and restoration. By goal, we consider the focus on the studies - for instance, whether a solution is focused on updating software, fixing a bug, vulnerabilities or patching specific attacks such as buffer overflow. Table~\ref{tab:instPatching} further shows whether a patching task is fully automated, needs manual input from humans to trigger an automated task (i.e., semi-automated) or fully depends on humans. In most of the instruction level patching solutions, patch deployment is followed by patch generation, which are done automatically~\cite{Inst5-chen2013safestack, xu2020automatic-Kernel4F-Inst2Vulmet, Inst1-Karmachen2017adaptive, Inst3-WordPatch-chamith2016living}. We have further identified the scope of the patching. Due to variations in technology and languages, a patching solution is not universal; thus, the scope of a particular study shows if the proposed approach is for Linux, a particular language like C or a cloud system. From Table~\ref{tab:instPatching}, we observe that most of the instruction-level patching approaches are automated and span among several systems and application domains. 

\begin{table}[thb]
    \centering
    \caption{Instruction-level patching}
    \begin{tabular}{p{2.6cm}|p{3.3cm}|p{2.5cm}|p{2.2cm}| p{2cm}}
    \hline
    \textbf{Study}   &  \textbf{Goal} & \textbf{Task perform} & \textbf{Task type } & \textbf{Scope}\\ 
     \hline
        D-linking \cite{inst8-DSU-Hics2005} & Updating software & Patch generation and deployment & Semi-automated & C-like language  \\
    \hline
        Safestack \cite{Inst5-chen2013safestack} & Buffer overflow attacks & Patch diagnosis, generation and deployment & Automated &Cloud system and data centre \\ 
     \hline
        Instaguard \cite{Inst4-chen2018instaguard} & Vulnerabilities in Android Program - integer overflow, buffer overflow, out-of-hour accesses and logic bugs& Patch generation and deployment & Semi-automated &  Android System (Binaries) (Nexus~5)\\
    \hline
        Replus \cite{Inst6-Function5-Replus-chen2015framework} & Security patches and long lived function & Patch generation and deployment & Semi-automated & Program in C language\\ 
     \hline 
        Vulmet \cite{xu2020automatic-Kernel4F-Inst2Vulmet} & Vulnerabilities in Android kernel & Patch generation and deployment & Automated & Mobile system \\
    \hline
        Dynsec \cite{Inst7-payer2013hotAsap, Inst7-payer2013dynsec} & Vulnerability in Application (e.g., Apache web server) & Patch generation and deployment & Semi-Automated & Networked system\\
    \hline
        KARMA \cite{Inst1-Karmachen2017adaptive} & Vulnerabilities in Android kernel & Patch integrity verification and deployment & Automated  & Android Kernel\\
    \hline 
        WordPatch \cite{Inst3-WordPatch-chamith2016living} & x86 architecture probe toggling & Patch deployment & Automated  & Processor \\
    \hline
    Kgraft \cite{web:kgraft-inst10} & Vulnerabilities & Patch deployment & Automated & Linux Kernel \\
    \hline
    \end{tabular}
    \label{tab:instPatching}
\end{table}

Instruction-level patching is performed by either inserting new instructions or modifying existing instructions. Chen et al.~\cite{Inst6-Function5-Replus-chen2015framework} have considered both scenarios. In one scenario extra jump instructions are added. In another scenario, patches modify existing instructions to enable boundary condition safety checks. As shown in Table~\ref{tab:instPatching}, a set of studies has used instruction level patching to mitigate software vulnerabilities~\cite{Inst5-chen2013safestack, Inst7-payer2013dynsec, Inst6-Function5-Replus-chen2015framework}. For instance, Chen et al.~\cite{Inst5-chen2013safestack} have proposed to monitor program instructions to check stack buffer overflow attacks. Here, instead of replacing a function, the designed patch application targets the specific instructions that trigger buffer overflows. They have proposed to automate the patch process from diagnosis to applying patches at runtime without interrupting the execution of the ongoing service. The proposed approach tracks function calls, however, it does not replace the whole functions and also does not consider cross-function changes. In another study, Chen et al. proposed to perform instruction-level patching for security patches that do not modify function argument types or amount~\cite{Inst6-Function5-Replus-chen2015framework}.  Arnold et al.~\cite{Function1KspliceArnold2009} have also proposed to patch individual instructions, however, they mainly replace entire functions. Payer et al.~\cite{Inst7-payer2013dynsec, Inst7-payer2013hotAsap} have proposed a novel approach to combine sandbox with a runtime patching approach to maintain the integrity and availability of a system. 

Similar to the above works, instruction level patching is also applied in Android OS~\cite{Inst1-Karmachen2017adaptive, Inst4-chen2018instaguard, xu2020automatic-Kernel4F-Inst2Vulmet} and Linux kernel~\cite{Function9-rommel2019multiverse}. In Instaguard, Chen et al.~\cite{Inst4-chen2018instaguard} have considered the function name to easily locate the memory addresses that trigger vulnerability exploitation, however, they do not replace the whole function with a patched version. They considered fine-grained line-oriented changes. In KARMA~\cite{Inst1-Karmachen2017adaptive}, the main target is to protect the integrity of the Android kernel whereas Instaguard protects the kernel from exploits originating from user space. Malicious inputs are filtered to prevent vulnerable kernel code from being exploited~\cite{Inst4-chen2018instaguard}. In another study, Xu et al.~\cite{xu2020automatic-Kernel4F-Inst2Vulmet} have emphasized on automatic generation of patches and proposed an approach, namely Vulmet, that automatically generates patches for Android kernel vulnerabilities. Vulmet automatically computes necessary memory location in a function and replaces the vulnerable code with the specified patch. Instruction-level patching techniques are also used for investigating the effects of thread safety issues and analyzing the reaction and visibility of other execution threads on modifying x86 code~\cite{Inst3-WordPatch-chamith2016living}.

\subsection{Function-level Patching} 

Function-level patching is coarser compared to instruction-level patching as it targets patching whole functions~\cite{Function1KspliceArnold2009, Fucntion2-AppWrapper-lee2020your}. For example, it replaces a buggy function with a patched one to eliminate vulnerabilities. A set of studies have proposed to perform function-level patching~\cite{ Function1KspliceArnold2009, Fucntion2-AppWrapper-lee2020your,  Function3-OSSPatcher-duan2019automating, Function4-hotpaccher-jeong2017functional, Inst6-Function5-Replus-chen2015framework, Function6-chen2007polus,Fucntion7-araujo2020JIT, Function8-Cure-zhao2016, Function9-rommel2019multiverse}. Performing function-level patching requires taking cross-function dependencies into consideration. In addition, long-living functions pose further challenges, as a suitable patching time must be determined. The steps of locating the patch target address are not much different compared to instruction-level patching. Table~\ref{tab:fucntionPatching} summarizes the studies related to function-level patching. It shows most of the function-level patching studies, patch generation is performed before patch deployment. Similar to instruction-level patching the scope of the studies span different domain, OS and application.  

\begin{table}[thb]
    \centering
    \caption{Function-level patching}
   \begin{tabular}{p{2.6cm}|p{3.5cm}|p{2.5cm}|p{2.2cm}| p{2.5cm}}
    \hline
    \textbf{Study } & \textbf{Goal} & \textbf{Task perform} & \textbf{Task type } & \textbf{Scope}\\ 
    \hline
         Kshot \cite{Function7Kshot} & Patching vulnerabilities & Patch generation and deployment & Semi-automated & Linux Kernel\\
    \hline
       Replus \cite{Inst6-Function5-Replus-chen2015framework} & Updating software & Patch generation and deployment & Automated & Program in C language\\
    \hline
    DUSC \cite{Function10-Orso2002DUSC} & Updating Java based Software & Patch generation and deployment & Automated, Semi-Automated & Java based application\\ \hline
      CURE \cite{Function8-Cure-zhao2016} & Generating safe patches and updating software  & Patch generation and deployment & Automated & - \\
    \hline
      POLUS \cite{Function6-chen2007polus} & Updating software & Patch generation and deployment & Semi-automated & Contemporary server software\\
    \hline
     Piston \cite{Function10-salls2017piston} & Vulnerabilities in the embedded device - stack-based buffer overflow, heap overflow & Patch generation and deployment & Automated, Semi-automated & Embedded device - remote patching\\
     \hline
      Ksplice \cite{Function1KspliceArnold2009} & Patching vulnerabilities & Patch deployment & Automated & Linux Kernel\\
      \hline
          Multiverse \cite{Function9-rommel2019multiverse} & Dynamic variability in performance critical paths & Patch deployment & Semi-automated & Cloud system, legacy code base   \\
    \hline
     AppWrapper \cite{Fucntion2-AppWrapper-lee2020your} & Applying dynamic security policies security functions & Patch deployment & Automated &  Mobile devices (Android)\\ \hline
      OSSPatcher  \cite{Function3-OSSPatcher-duan2019automating} & Patching Vulnerabilities & Patch generation and deployment & Automated & Vulnerable mobile applications\\
      \hline
     Hotpaccher \cite{Function4-hotpaccher-jeong2017functional} & Updating software & Patch deployment & Automated & ELF binary application\\ \hline
   C-MultiVerse \cite{function11rothberg2016CompMultiverse} & Dynamic variability in system software & Patch generation and deployment& Automated & Cloud system, legacy code base \\\hline
  Gitar \cite{Functiona12ruckebusch2016gitar} & Updating network stack  of constrained devices  & Patch deployment & Automated & IoT, M2M, constrained devices' single Rime modules  \\ \hline
   BinPatch \cite{Function13hu2019BINPATCH} & Patching Vulnerabilities in binary program (file level) & Patch generation, deployment and restoration & Automated & Binary code \\ \hline
   UpdateCalculus \cite{Function14-bierman2003UpdateCalculus} & Formally Verifying patches before deployment & Patch verification & Manual & Theoretical\\ \hline
    \end{tabular}
    \label{tab:fucntionPatching}
    \vspace{-10pt}
\end{table}

Several of the runtime patchings approaches at function-level are designed for security patches~\cite{Function1KspliceArnold2009, Fucntion2-AppWrapper-lee2020your, Function3-OSSPatcher-duan2019automating, Function13hu2019BINPATCH}. For instance, Duan et al.~\cite{Function3-OSSPatcher-duan2019automating} have proposed OSSPatcher that automatically identifies vulnerable functions, generates binary patches and performs patch injection at runtime. Similarly, Lee et al.~\cite{Fucntion2-AppWrapper-lee2020your} have proposed an Appwrapper toolkit to inject additional security code on a per-method (i.e., function) basis into insecure apps enabling the use of dynamic policies to enhance overall application security. Function-level patching is also performed to replace a whole vulnerable function by linking it with a new function or replacement code into the kernel. The proposed method is considered very useful for large server environments that are highly utilized by multiple users~\cite{Function1KspliceArnold2009}. Research is also seen in patching vulnerabilities of binary programs via code transfer and binary rewriting in functions~\cite{Function13hu2019BINPATCH}. Table~\ref{tab:fucntionPatching} shows most of the function level patching is performed automatically without user intervention.

Existing works on patching kernel at runtime rely on the host Kernel and consider the Kernel to be always trusted. However, Zhou et al.~\cite{ Function7Kshot} have emphasized the fact that Kernel can be malicious and untrusted. Thus, they have proposed a reliable kernel runtime patching framework (albeit at function granularity) leveraging Trusted Execution Environments (TEE) that is hardware-assisted to prepare and deploy kernel patches. The proposed approach Kshot, prepares and deploys patches that do not need a trustworthy kernel patching mechanism~\cite{ Function7Kshot}. 

Considering the limited capacity of the embedded devices (e.g., lack of update functionality), Salls et al.~\cite{Function10-salls2017piston} have proposed an approach, Piston, to perform remote patching on embedded devices. Piston mainly performs patching at the function level. It is designed to force patches onto a vulnerable system by exploiting the vulnerabilities and taking control of the vulnerable process. For some vulnerabilities like stack-based buffer overflows, it supports automated patching whereas for others it is semi-automated and requires input from the analyst. Similarly, Rucjerbusch et al.~\cite{Functiona12ruckebusch2016gitar} have proposed a tool Gitar to enable runtime patching of applications in OSes running on IoT, M2M and resource-constrained devices.  Table \ref{tab:fucntionPatching} shows that function-level runtime patching approaches are also adopted to perform dynamic variability in software systems~\cite{Function9-rommel2019multiverse, function11rothberg2016CompMultiverse}. Most of these approaches are compiler assisted and implement a function multiverse in the compiler to perform binary patching. 

Function-level patching is noticed to perform runtime patching at the production level, especially for C program ~\cite{Inst6-Function5-Replus-chen2015framework}. The authors have proposed a framework, Replus, to build environment-aware patches by separating the responsibilities of developers and customers where patch generation is performed in the developer environment and deploying or applying the patch is performed in a customer environment. The main focus of Replus is to provide a practical and efficient runtime patching system that does not need compiler support. Different from Replus, Orso et al.\cite{Function10-Orso2002DUSC} have proposed a tool, DUSC, to perform runtime patching in Java-based applications.

\subsection{Process-level Patching}

Process-level patching aims to replace the whole process with a new one that contains a fixed or updated functionality~\cite{Process1-mugarza2020cetratus, Process3-giuffrida2013safe-Proteos, Process8-hayden2012kitsune, Process7-ramaswamy2010katana}. In contrast to function-level patching, updating the whole process might be easier in some cases as only OS dependencies and resources need to be tracked. For instance, files and network sockets need to be detached from the old process and transferred to the new process~\cite{Process4-rommel2019waitFree}. The main complication in this area is maintaining the internal resource states such as current file pointers and network protocol states. Table~\ref{tab:ProcessPatching}
shows the key studies that have considered runtime patching at the process level. 
Among the existing studies, Rommel et al.~ \cite{Process4-rommel2019waitFree} have proposed to move individual threads to new address space through predefined quiescent states. The proposed approach attempts to minimize waiting time by preparing the clone of the address space as opposed to the stop-the-world approach.

\begin{table}[htb]
    \centering
    \caption{Process-level  patching}
    \begin{tabular}{p{2.5cm}|p{3.3cm}|p{2.5cm}|p{2.2cm}| p{2cm}}
    \hline
    \textbf{Study } & \textbf{Goal} & \textbf{Task perform} & \textbf{Task type } & \textbf{Scope} \\ 
    \hline
 Cetratus \cite{Process1-mugarza2020cetratus} & Updating software securely & Patch deployment & Semi-automated & Safety-critical systems\\ \hline

 Proteos \cite{Process3-giuffrida2013safe-Proteos}  & Updating software safely and stably & Patch deployment & Semi-automated & Custom OS (Minix based)\\ \hline
  WaitFree \cite{Process4-rommel2019waitFree} & High performance, Multi-threading support, Read-only memory segments & Patch deployment & Semi-automated & Linux on AMD64 \\ \hline
  MVEDSUA \cite{Process5-pina2019mvedsua}  & Updating software & Patch deployment and testing and rollback & Semi-automated & C-based application\\ \hline
 TEDSUTO  \cite{Process6-pina2016tedsuto}  & Updating software safely & Patch deployment and verification & Semi-automated & Java-based system\\ \hline
   Katana \cite{Process7-ramaswamy2010katana} & Updating critical security or functionality& Patch generation, deployment & Automated &  ELF binaries (Linux)\\ \hline
  Kitsune \cite{Process8-hayden2012kitsune, Process8-hayden2014kitsune}     & Updating software & Patch deployment and rollback & Semi-automated & C-based system\\ \hline
     Ginseng \cite{Process9-neamtiu2006Ginseng} & Updating software safely & Patch generation and deployment & Semi-automated & C programs \\ \hline
    \end{tabular}
    \label{tab:ProcessPatching}
    \vspace{-5pt}
\end{table}

As shown in Table~\ref{tab:ProcessPatching}, we observe that most of the process-level patching solutions are semi-automated; hence, actual patch deployment requires human assistance. Process-level patching studies focus mainly on patch deployment rather than patch generation. The scope of the process-level patching solution typically targets a specific OS or programming language. A common way to achieve process-level patching is the usage of multi-version execution techniques~\cite{Process8-hayden2012kitsune, Process8-hayden2014kitsune}. Both update-safe points and corresponding data transformation functions must be predefined by the developers to make runtime updating possible. One of the proposed approaches, Kitsune~\cite{Process8-hayden2012kitsune, Process8-hayden2014kitsune}, implements single- and multi-threaded C-based application updating. The authors propose a novel tool to assist software developers in generating state transformation and transfer routines. Some light annotation is, however, still required from the developers. MVEDSUA~\cite{Process5-pina2019mvedsua} is an extension of Kitsune~\cite{Process8-hayden2012kitsune} with the use of N-version execution framework Varan\footnote{Varan \cite{hosek2015varan}, is a rich framework in terms of applicability and is particularly useful for software updating purposes. The initial motivation behind the proposed approach is to run multiple copies of the same code with various modifications to minimize potential vulnerability risks. For instance, transparent failover can be achieved if only one of the copies is exploited. In the context of software updating, the same principle can be used to detect behavior deviations caused by the updated code as, for example, done in~\cite{Process5-pina2019mvedsua}.}~\cite{hosek2015varan}. The key insight behind MVEDSUA is the ability to patch a second copy of the code, which runs simultaneously with the original code. Subsequent checks of old and new code behavior make it possible to verify and roll back broken patches if any discrepancies are detected. This approach requires operator assistance to define the potentially expected behavior discrepancies explicitly.

Several of the process-level patching approaches focus on the security and safety of the system after applying patches~\cite{Process1-mugarza2020cetratus, Process3-giuffrida2013safe-Proteos, pina2016tedsuto}. For instance, Giuffirida et al.\cite{Process3-giuffrida2013safe-Proteos} have focused on safe and predictable updates of OS states. The proposed solution, Proteos, performs transactional live updates at the process level and requires programmer-provided state filters to limit updates to certain points in time only. Similarly, research is noticed on the consideration of the safety and security of a system while deploying patches~\cite{Process1-mugarza2020cetratus} where  a tool, Cetratus, is proposed for performing and verifying runtime patches in safety-critical systems. Pita et al.~\cite{pina2016tedsuto} have proposed TEDSUTO, which is focused specifically on patch verification steps. Old and new code behaviors are systematically tested, and discrepancies are identified as potential patch-related errors. This approach enables safe patch testing prior to applying it in a production environment. Similar to MVEDSUA~\cite{Process5-pina2019mvedsua}, intended behavior changes must be explicitly annotated by developers.

Ramaswamy et  al. have proposed Katana \cite{Process7-ramaswamy2010katana} which attempts to generate patches from source changes and safely apply them at run time. An interesting feature of the proposed solution is automatic safe update point detection. Specifically, the process execution stack is constantly monitored for this purpose, with suitable execution points being deduced on the fly. Stop-the-world technique is applied when a safe update point is encountered to replace the old code with the patched version. Authors, however, admit that their approach does not guarantee time bounds as a suitable execution point may not necessarily be reachable. Similar to Katana \cite{Process7-ramaswamy2010katana}, Neamtiu proposed a tool, Ginseng \cite{Process9-neamtiu2006Ginseng}, that aims to integrate patch generation and deployment steps together. Function indirection and type wrapper are the key techniques used by Ginseng. Additional annotations are required from the developers to define safe update points. 

Different from existing approaches, Bierman et al.~\cite{Function14-bierman2003UpdateCalculus} have focused on formal foundation and conceptual reasoning about safe updating. Namely, an \textit{update calculus} is proposed to enable formal reasoning about various code properties such as type safety or semantic correctness. Having a strong theoretical focus, this work does not tackle technical implementation-related issues such as specific programming languages or OS.

\subsection{Container-level Patching}

Containerization is an emerging technology that has become widely popular in distributed computing applications. With the increasing adoption, various attacks, vulnerabilities and security challenges are noticed in distributed computing environments such as data centers and the cloud. Patching vulnerabilities in the container is not straightforward~\cite{Container1-Opatch-tunde2020, Container2-tunde2020selfPatch, Container3-sun2020blockchain,  Container4-ayres2021continuous}. Existing patching techniques often are not suitable for containers due to their short life and ephemeral nature~\cite{Container1-Opatch-tunde2020, Container2-tunde2020selfPatch}. Tunde-Onadele et al. have proposed lightweight on-demand patching techniques to patch containerized applications at runtime~\cite{Container1-Opatch-tunde2020, Container2-tunde2020selfPatch} considering the resource constraints. Research in container-level patching techniques is noticed to enhance cloud security where blockchain-based integrity checkers are proposed to detect vulnerabilities in container images~\cite{Container3-sun2020blockchain}. The proposed approach further aids in generating and dynamically replacing vulnerable instances. 


\begin{table}[thb]
    \centering
    \caption{Container-level, VM-level, Hypervisor-level and Kernel-level patching}
    \begin{tabular}{p{2.9cm}|p{3.5cm}|p{3.5cm}|p{1.5cm}|p{2cm}}
    \hline
    \textbf{Study}   &  \textbf{Goal} & \textbf{Task performed} & \textbf{Task type } & \textbf{Scope} \\ 
    \hline
      \rowcolor{light-gray}\multicolumn{5}{c}{\textbf{Container}} \\ \hline
     
      \hline
      SelfPatch \cite{Container2-tunde2020selfPatch, Container1-Opatch-tunde2020} & Patching vulnerabilities in container image& Patch deployment & Automated & Server \\ \hline
      BlockChainPatch  \cite{Container3-sun2020blockchain} & Patching vulnerability, malware (repository level) & Patch deployment & Automated & Container image security\\
      \hline
      ContainerECU \cite{Container4-ayres2021continuous} & Patching vulnerabilities, bugs and errors & Patch deployment & Automated & Automotive software\\
      \hline
      \rowcolor{light-gray}\multicolumn{5}{c}{\textbf{VM}}\\ \hline
        Shadow Patching \cite{VM1-le2014shadowPatching} & Patching vulnerabilities & Patch deployment and testing & Automated & Cloud system and data centre \\
    \hline
        vPatcher \cite{VM2-zhang2014vpatcher} & Patching vulnerabilities & Patch deployment & Automated& Cloud platform and data centre\\
    \hline
     \rowcolor{light-gray}\multicolumn{5}{c}{\textbf{Hypervisor}}\\ 
     \hline
        Orthus \cite{hypervisor2-Orthus-zhang2019fast} & Patching vulnerabilities and feature updates & Patch deployment & Automated  & Cloud system (Alibaba cloud)\\
    \hline
        HyperFresh \cite{hypervisor1-hyperFresh-doddamani2019} & Patching vulnerabilities and feature updates & Patch deployment & Automated & Cloud system and
data centre\\
  \hline
    \rowcolor{light-gray}\multicolumn{5}{c}{\textbf{Kerne}l}\\
     \hline
        Seamless \cite{Kernel3Seamless} & Vulnerabilities and bugs& Patch deployment & Automated & Linux kernel\\
    \hline
        Kup \cite{Kernel1Kup} & Vulnerabilities, bugs, feature updates & Patch deployment and testing & Automated& Linux kernel\\ \hline
     CRIU \cite{kernel3Crui} & Upgrading full kernel & Patch deployment & Semi-automated & Linux kernel\\ \hline
   
    \end{tabular}
    \label{tab:kernelPatching}
\end{table}

Recent studies have leveraged container virtualization and lightweight features to perform runtime patching in the automotive domain. For example, Ayres et al.~\cite{Container4-ayres2021continuous} have proposed to utilize the virtualization features of container technology in the Electronic Control Unit (ECU) of automotive vehicles to perform patching at runtime. The proposed approach is suitable for automotive functions which do not involve vehicle operation or safety and is instead focused on patching software related to autonomous driving or subsystems such as climate control. Table \ref{tab:kernelPatching} shows that most of the container-level patching approaches are focused on automatically patching vulnerabilities, bugs or errors. Therefore, deploying patching at runtime is the main task of the studies categorized in container-level patching. However, the studies span server systems, container image security and automotive software.
Table~\ref{tab:kernelPatching} further summarizes the studies that proposed approaches for VM, Hypervisor and Kernel level patching.

\subsection{VM-level Patching}

VM-level patching granularity is conceptually similar to process-level, with the main difference relating to technical aspects. For instance, while replacing an existing process requires interacting with the OS, replacing a virtual machine would require interacting with the virtualization hypervisor\footnote{Hypervisors support the creation and running of VMs, which is also known as Virtual Machine Monitor (VMM).}. We find only a few studies where a patch is performed at VM-level~\cite{VM1-le2014shadowPatching, VM2-zhang2014vpatcher}. Table~\ref{tab:kernelPatching} shows that VM-level studies mainly perform patch deployment automatically to patch vulnerabilities in cloud systems and data centers.
Among them, Le et al.~\cite{VM1-le2014shadowPatching} have proposed a shadow patching technique to reduce the downtime caused by patching VM running on the managed environment such as data centers. First, the proposed approach creates a replica of the VM that needs to be patched. Then, it deploys the patches in the replica VM, performs testing and makes necessary changes before applying the patch to the original VM. Finally, when the patches are applied to the original VM, the running application and data are merged from the replica VM to the original VM. This last step is similar to the VM teleport technique where current workloads are seamlessly transferred to the new VM instance\footnote{\url{https://docs.oracle.com/en/virtualization/virtualbox/6.0/admin/teleporting.html}}.

Different from the shadow patching approach, Zhang et al.~\cite{VM2-zhang2014vpatcher} have utilized VM introspective capabilities to monitor the behavior of VM using hypervisors to perform patching at runtime. Furthermore, they aim to perform data patching transparently in VM using hypervisor assistance. The proposed approach, vPatcher, is deployed outside the target guest VM to intercept and scan VM network traffic based on a vulnerability signature. This allows to filter out the network connections based on the vulnerability signatures.

\subsection{Hypervisor-level Patching}

By hypervisor-level runtime patching, we consider studies that aim to replace the whole virtualization hypervisor without or minimally affecting the underlying virtual machines. Patching a hypervisor involves changing a large amount of dynamic data structures and is significantly complicated by the necessity to keep multiple underlying guest kernels intact and in sync with the hardware state. We observe a set of studies where hypervisors are patched at runtime without disrupting the running VMs~\cite{hypervisor1-hyperFresh-doddamani2019, hypervisor2-Orthus-zhang2019fast}. Table~\ref{tab:kernelPatching} shows that hypervisor-level runtime patching is commonly applied to patch vulnerable cloud infrastructure without human involvement which mainly occurs in cloud and data center environments. In contrast to VM migration, hypervisor or VMM (Virtual Machine Manager) patching is typically performed in the same physical node and does not introduce significant network traffic. Thus, replacing or patching a hypervisor achieves better performance and lower downtime than VM live migration. 

One technique to patch hypervisor at runtime is to copy the whole state of all active VMs from the running hypervisor to a new hypervisor (upgraded)~\cite{hypervisor2-Orthus-zhang2019fast}. The proposed techniques aim to seamlessly pass through devices to the new hypervisor automatically without losing any ongoing activities or operations state. Similarly, Doddamani et al.~\cite{hypervisor1-hyperFresh-doddamani2019} have proposed to transparently replace a hypervisor with a new one without disrupting or shutting down the VMs. The proposed approach, namely HyperFresh, introduces an intermediate layer to perform live hypervisor replacement. This layer maps the memory of the running VMs to the new locations belonging to the new hypervisor located in the same physical machine. These approaches focus on patching vulnerabilities and adding new features of cloud systems without interrupting the running VMs or shutting down the serves. While technically complicated, hypervisor-level patching is crucial in large commercial cloud deployments. Table~\ref{tab:kernelPatching} also shows that the studies on Hypervisor-level patching are mainly focused on deploying patches at runtime for patching vulnerabilities or updating features at cloud systems and data centers.

\subsection{Kernel-level Patching}

We consider the patching solutions where a whole kernel is switched to a new kernel to apply all kinds of patches without having to restart the kernel~\cite{Kernel1Kup} as Kernel-level runtime patching. We noticed a few studies \cite{Kernel1Kup, Kernel3Seamless, kernel3Crui} that have performed runtime patching at kernel-level. Kernel-level patching is executed through the use of kernel space and user space checkpoint and restart mechanism~\cite{Kernel1Kup, Kernel3Seamless, kernel3Crui}. Table~\ref{tab:kernelPatching} also shows that few studies have proposed to replace the whole Kernel while patching at runtime. Unsurprisingly, from Table~\ref{tab:kernelPatching} we can see that the kernel-oriented solutions mainly target Linux, most likely due to the source code availability. The goal of these studies includes patching vulnerabilities, fixing bugs, updating features and upgrading an entire kernel. 

In place kernel switch is a popular technique to update OS that heavily modifies various subcomponents of the kernel~\cite{Kernel3Seamless}. The proposed approach mainly targets restoring the applications entirely in kernel space. After applying patches to the Kernel, it is essential to ensure that the changes do not break or corrupt the existing functionalities and applications. Thus, the testing process of kernel-level patching is costly and tedious~\cite{xu2020automatic-Kernel4F-Inst2Vulmet}. Kashyap et al.~\cite{Kernel1Kup} have proposed to perform runtime kernel patching without modifying the kernel source code. They have adopted binary patching techniques, userspace checkpoint and restore, and optimized in-place Kernel switching to perform the kernel patching. Alternatively, in order to simplify kernel updating, resource (such as running processes and containers) handover assistance can be utilized~\cite{kernel3Crui}. For instance, a widely used tool, CRIU\footnote{\url{https://www.criu.org/Main_Page}} makes it possible to freeze critical processes, replace the whole old Kernel with a patched/new version and restore the frozen processes. While CRIU was initially designed for live container migration, further project development has enabled the runtime kernel upgrading scenario.

In summary, we observe that function-level patching techniques are widely proposed in the literature, followed by instruction-level and process-level patching. The key reason we notice is the granularity of the patch; with function and instruction level patching, there require more minor changes in the resources as shown in Figure~\ref{fig:patchlevel}. On the other hand, we notice little research in patching heavy loaded resources such as VM, container, hypervisor and Kernel due to the complexity and dependency of the resources and tasks with other resources. 

\section{Patch Deployment strategies, workflow and Capabilities} \label{sec:patchStrategiesCapability}

This section provides an overview of the patch strategies the reviewed patching solutions employed, along with patch deployment workflows commonly implemented in practice.  \textit{State transformation} and\textit{ co-exists \& decay} are two identified patch strategies that are discussed in detail in Section \ref{subsec:strategy}. Despite the high variability of technical implementations, a detailed look into the patch deployment steps in Figure \ref{fig:life_cycle} reveals several commonalities between the existing runtime patching solutions. Individual patch deployment steps vary in terms of patch strategies which are discussed in Section~\ref{subsec:patcWorkflow}. Crucial technical capabilities exhibited by the reviewed solutions are discussed in Section~\ref{subsec:patchCapability}.

\subsection{Runtime Patching Strategies}\label{subsec:strategy}

The analysis of existing runtime patching approaches reveals two main patch deployment strategies (i) State transformation and (ii) Co-exist \& decay. These patch strategies refer to the conceptual models underpinning the patch handling steps. Here, we discuss the general goals and trade-offs of the identified patching strategies. 

\subsubsection{\textbf{State Transformation}}

\begin{figure}[htb]
     \centering
     \begin{subfigure}[b]{0.6\textwidth}
         \includegraphics[width=\textwidth]{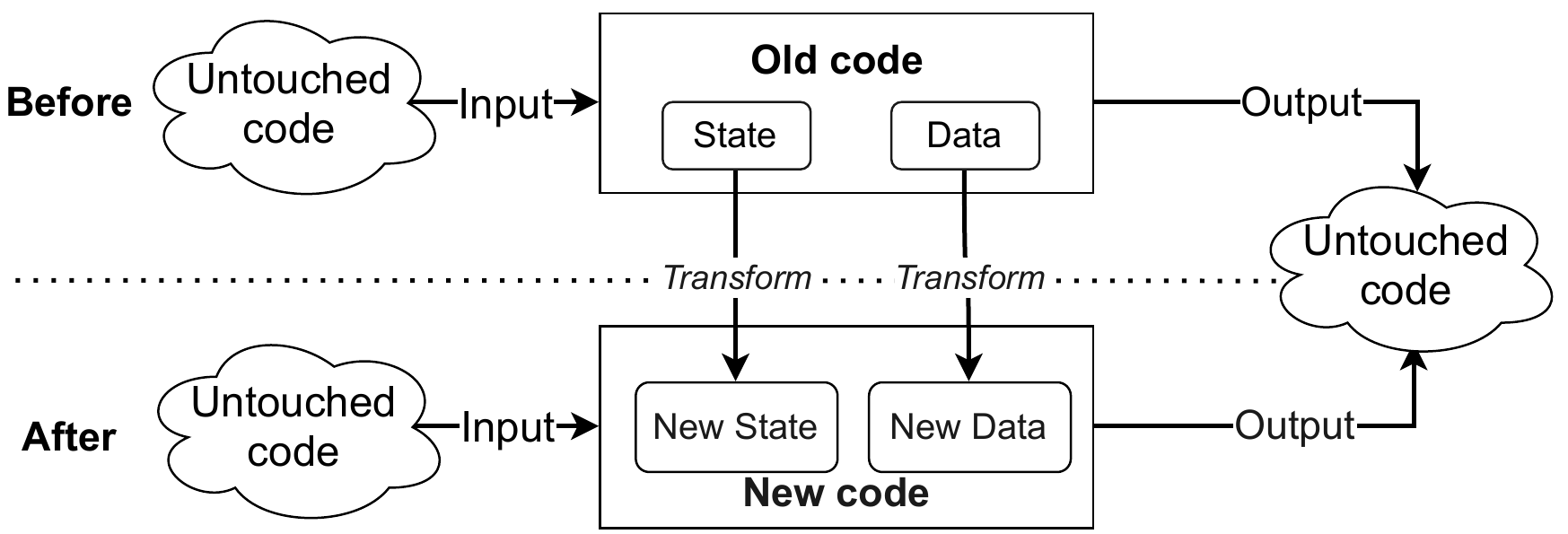}
         \caption{State transformation}
         \label{fig:StateTransformation}
     \end{subfigure}
     \vfill
     \begin{subfigure}[b]{0.8\textwidth}
         \centering
          \centering
         \includegraphics[width=\textwidth]{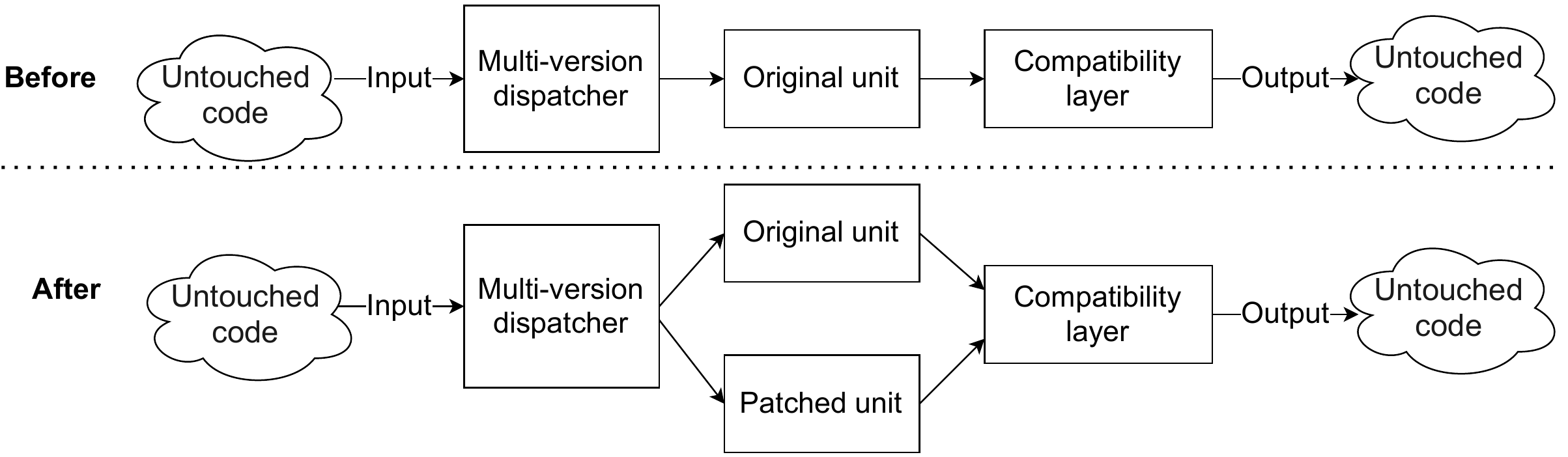}
         \caption{Co-exists \& decay}
         \label{fig:CoexistsDecay}
       
     \end{subfigure}
        \caption{Conceptual view of patch strategy (a) co-exists \& decay and (b) state transformation before and after applying a patch}
        \label{Fig:Patchstrategy}
\end{figure}

State transformation attempts to modify the currently running code to fix or remove the unwanted behavior. State transformation line of thought aims to convert the old code along with all the necessary resources, data or states to match the expected behavior of the patched code as shown in Figure \ref{fig:StateTransformation}. This approach ensures that the patched code is effective immediately as soon as the state or data transformation is completed \cite{Inst4-chen2018instaguard, Kernel1Kup, Kernel3Seamless}. State transformation strategy can be implemented in cooperative and uncooperative manners. 

 \textit{Cooperative} patching relies on some degree of assistance from the old code. Usual assistance avenues include auxiliary compiler- or developer-provided code or annotations that simplifies determining quiescent execution state and/ or data transformation procedures. This, however, requires designing the application to provide such cooperation from the grounds up and is not suitable for legacy systems. Hence, cooperative state transformation is mainly viewed as part of new development practices and tooling.

\textit{Uncooperative} patching implies that the old code is completely unaware of possible updates, and the whole patching burden falls on the patching subsystem. In particular, this means that the patching code has to possess intimate knowledge of the internals of the old code. Existing resources must be located, detached from the old code and converted. Depending on the type of software, this would typically mean parsing and modifying old code RAM contents. While this type of patching is technically more challenging than the cooperative one, the main advantage is the broader applicability. Uncooperative patches can be potentially applied to a wider set of existing legacy software systems.

The main obstacles encountered by state transformation strategies are that the transformation process itself is not instantaneous and not applicable at any moment in time. Modifying arbitrary data bytes in memory may negatively impact current execution threads that actively use modified data. Furthermore, uncoordinated data modifications may introduce inconsistencies between different parts of code, which can lead to unforeseen consequences, side effects and behavior. Thus, state transformation strategy involves waiting for a less active (quiescent) state to be reached for software or code execution. Reaching a quiescent state allows for safely transforming the necessary memory objects. However, this is not straightforward as some code may include a long-living function that rarely (if ever) ceases execution.
Furthermore, due to external factors, highly-loaded applications may never reach a quiescent (safe-to-update) state. In addition, as state transformation is not instantaneous, any concurrent access to the memory object being transformed must be prevented. Therefore, code execution is typically briefly suspended while the transformation is conducted. Note that while such execution suspension is technically a service interruption, short interruptions are assumed to go unnoticed by external observers. 
State transformation attempts to modify the currently running code to fix or remove the unwanted behavior. State transformation line of thought aims to convert the old code along with all the necessary resources, data or states to match the expected behavior of the patched code as shown in Figure~\ref{fig:StateTransformation}. This approach ensures that the patched code is effective immediately as soon as the state or data transformation is completed~\cite{Inst4-chen2018instaguard, Kernel1Kup, Kernel3Seamless}.

\subsubsection{\textbf{Co-exist \& Decay}}

Arbitrary state transformations may be either prohibitively computationally expensive or impossible for complicated software commonly used in practice~\cite{gupta1996formal}. Thus, an alternative strategy \textit{Co-exist \& decay} is proposed that attempts to side-step the transformation-related issues by exploiting the transaction-based nature common to modern software services.  Depending on the granularity in consideration, such transactions can vary from individual CPU instructions and functions to higher-order logically independent user-software interactions. Figure~\ref{fig:CoexistsDecay} shows a high-level overview of co-exist \& decay approach. Certain network-based software and web applications in particular are prime examples, where users periodically establish independent sessions or send network requests based on their needs.

Coexist \& decay strategy attempts to completely avoid any interruptions at the expense of a potentially more prolonged code activation wait period. Using various forms of indirection in code execution dispatching mechanisms enables having multiple code paths available. In coexist \& decay, old and new codes coexist in the system simultaneously. Any new external interactions go through the new code execution path, while currently, in-flight interactions are processed by the old code. As already started interactions finish off, old code execution paths are decayed (i.e., removed with associated memory objects disposal).  For instance, web-based applications benefit from the stateless nature of the underlying HTTP protocol, with the patches being possible to apply between requests. In addition, the time intervals between separate requests provide natural, safe update points. These features make web applications a perfect fit for coexist \& decay solutions.

In addition to enabling the use of multiple concurrent versions of code, another important role of the dispatcher is to perform actual code path selection at run time. Significant changes implemented by complex patches may introduce some incompatibilities, primarily with external code. For instance, patches that alter network protocols on the server side may render remote network clients (that are not yet updated accordingly) incompatible with the new protocol version. Thus, the compatibility layer of the dispatcher, as shown in Figure \ref{fig:CoexistsDecay} needs to differentiate between transactions/interactions on the basis of code version compatibility. Further, throughout execution, once all existing interactions for a given version are finished, the associated code path can be decayed.

\subsection{Runtime Patching Deployment Workflow}\label{subsec:patcWorkflow}

\begin{figure}[hbt]
    \centering
    \includegraphics[scale=0.4]{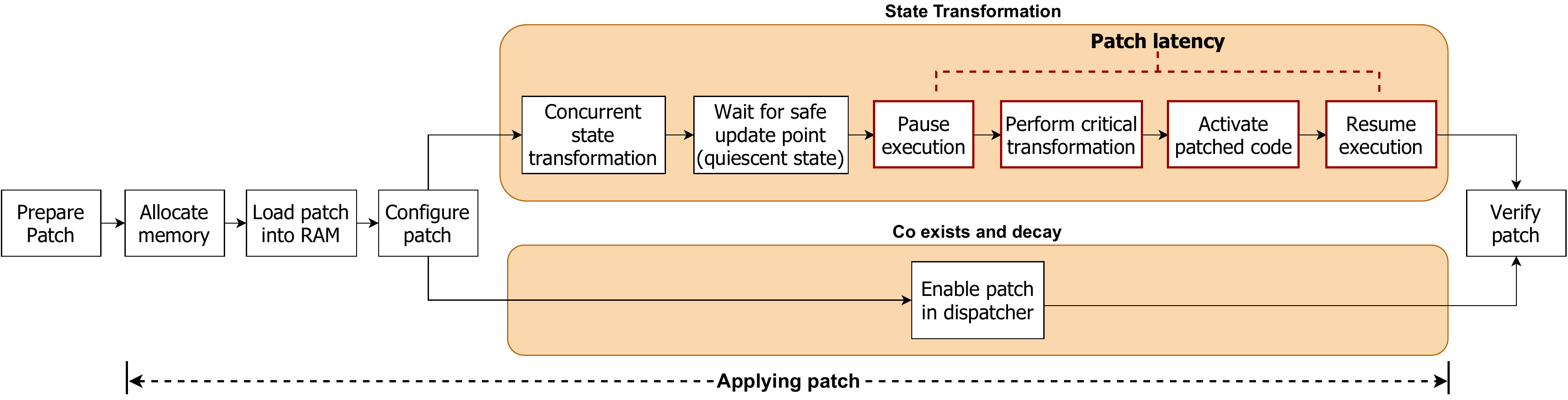}
    \caption{Workflow of deploying patching at runtime }
    \label{fig:runtimeWorkflow}
\end{figure}

Deploying or applying a patch at runtime consists of several preparatory steps followed by actual patch code activation steps. The preparatory steps of patch deployment are common for all patch approaches, while patch activation steps depend on the employed/ selected strategy (i.e., state transformation and co-exist $\&$ decay). As can be seen from Figure~\ref{fig:runtimeWorkflow}, allocating memory, loading patch into RAM and configuring patch are common steps. Supported software environments such as OS, programming language and machine architecture greatly affect how each of these steps are implemented by a given patching solution~\cite{Function7Kshot,Inst5-chen2013safestack,xu2020automatic-Kernel4F-Inst2Vulmet, Container1-Opatch-tunde2020}. Note that Figure~\ref{fig:runtimeWorkflow} outlines the conceptual steps that do not depend on the particular granularity chosen. The outlined steps need to be taken regardless of the granularity selected; however, the actual implementation may differ. The type of resources to transform or dispatch depends on the patch granularity and nature (Figure~\ref{fig:patchlevel}). For instance, patching a whole VM compared to patching an individual instruction typically requires transforming more resources (e.g., network sockets, file handles, hypervisor objects, etc.).

\textit{Allocating memory} to accommodate the patched code in RAM is a necessary step for all solutions except those that perform limited instruction overwriting in-place (such as~\cite{Inst3-WordPatch-chamith2016living}). \textit{Loading the patched code into RAM} is necessary regardless of the technical differences in the implementation. Lastly, \textit{configuration of a patch} may include sub-tasks such as decompression, decryption or digital signature verification than just copying patch bytes into memory. These sub tasks are optional and mainly implemented to increase performance or reliability. In some cases, the three preparatory steps may become completely unnecessary, complicated, impossible or require external intervention. For example, resource-limited embedded IoT devices may not have enough unused memory to fit the patched code into RAM, leading to the necessity to completely erase and re-flash the device through external means/devices.

Figure \ref{fig:runtimeWorkflow} shows further steps of \textit{patch activation} that depend on the general patching strategy employed. In efforts to minimize any potential system or service downtime, runtime patching solutions typically attempt to perform as many tasks as possible prior to activation of the patched code. For instance, \textit{concurrent state transformation} may occur with execution of old code wherever possible. As a consequence, once the \textit{quiescent state} is reached, the number of actual code activation steps would be reduced and thus take less time to complete. The \textit{patch code activation} step is time-critical as the \textit{patch latency} period between code execution \textit{pausing} and \textit{resuming} should be kept minimal. Thus, only \textit{critical state transformation} (that were not possible outside of quiescent state) should be conducted while the code execution is paused. While not having issues with quiescence, the patch enabling step in dispatcher used by coexist \& decay (as shown in Figure~\ref{fig:runtimeWorkflow}) must deal with non-instantaneous patch activation instead.

In both cases, the potential downside is that the old (potentially vulnerable) code would operate for some time, while the preparatory steps are taken, the quiescent state is not yet reached or new interaction starts. Having the vulnerable code still active, might pose a significant security risk for large-scale patches (i.e.,~VM- or kernel-sized), leading to the necessity to completely stop any system operation temporarily. The overall state transformation can then be performed in a safe manner. Such execution suspension would essentially be equivalent to traditional offline patching in terms of user experience disruption. 

\subsection{Patch Implementation Capabilities} \label{subsec:patchCapability}

Several generic capabilities implemented by existing patching systems have been identified throughout this study. In line with the presented patching workflow of Figure~\ref{fig:runtimeWorkflow}, we identify four main capabilities (see Fig \ref{fig:Patchtaxonomy}) - (i)~memory management, (ii)~resource transformation, (iii)~safe execution state determination and (iv)~execution path dispatching. The implementation details of the runtime patching systems exhibit these capabilities differently in the sense of limited applicability to certain use cases. For instance, in most of the reviewed solutions, these capabilities are limited to specific programming languages, OS-es and environments. Table~\ref{tab:study-capabilityl} summarizes an overview of the
patch implementation capabilities of different patch approaches along with execution or operational environment. It also shows the advantages or quality requirements promised to be fulfilled by a particular study.

\subsubsection{\textbf{Memory Management}} 
In contrast to traditional offline patching, which deals with file-level abstraction, one of the key tasks that a runtime patching system must solve is related to memory management. This includes finding necessary RAM locations, allocating memory for new code, relocating memory objects and so on \cite{Inst5-chen2013safestack, Function10-salls2017piston, hypervisor1-hyperFresh-doddamani2019,  Function7Kshot}. \textit{Determining RAM addresses} is typically required for lower-level patch granularities such as instruction- or function-level. More specifically, locating the old code address in memory is required to disable or overwrite vulnerable portions with the patched version. In some cases, such as when the length of the patched code is longer than the original, a new (free) memory region needs to be allocated and used. 

One of the popular solutions in these cases is the use of trampolines, where a few instructions of the original code are replaced to point to the patch code that is stored in the newly-allocated memory space~\cite{ Function7Kshot, Function10-salls2017piston, xu2020automatic-Kernel4F-Inst2Vulmet}. However, such trampolines do not address internal jumps and breaches to intermediate levels. This leads to the use of function trampolines during function-level patching. For example, Salls et al.~\cite{Function10-salls2017piston} have proposed Piston that replaces individual functions. Piston adopts trampolines that are essentially direct jump instructions at the beginning of the old functions to perform function replacement. In addition, implementing trampolines requires checking if the trampoline jump instructions fit entirely within the function body. This is partially mitigated in modern operating systems and compilers by aligning functions to 8-byte address boundaries, which should be enough for direct jump instructions.

Some patching systems attempt to simplify or completely eliminate the memory address locations task. For instance, relying on APIs exposed by existing dispatching mechanisms in the code, the patching system may not even need to find a concrete memory address to modify. Note that using OS-specific features such as Address Space Layout Randomization (ASLR) technology may significantly complicate address calculations~\cite{gu2016derandomizing}. While technically involved and situation-specific, locating memory addresses has known OS-specific solutions~\cite{gu2016derandomizing}.

Despite being highly platform-dependent, allocating additional memory regions for new code and data objects is relatively simple compared to specific RAM addresses finding tasks. The main problem is the amount of RAM to allocate. This might not be an issue for modern desktops and servers but could pose challenges in embedded and resource-constrained devices.

\begin{table}[hbt]
    \centering
    \small
    \caption{Runtime patching approaches with respective capabilities, granularities and quality requirements. The studies are mapped with four supported capabilities: Memory management (MM), Resource transformation (RT), Safe Execution State Transformation (SESD), and Execution Path Dispatching (EPD)}
    \begin{tabular}{p{1.95cm}|p{2.4cm}|p{0.4cm}|p{0.26cm}|p{0.47cm}|p{0.38cm}|p{1.35cm}|p{5cm}}
    \hline 
        \textbf{Study} & \textbf{Execution or operational environment } & \textbf{MM}&\textbf{RT} & \textbf{SESD}  & \textbf{EPD} & \textbf{Granularity}&\textbf{Focus (Quality Requirement)}   \\ \hline
      
      D-linking  \cite{inst8-DSU-Hics2005} & TALx86 & - & - & \checkmark & \checkmark & Instruction & Flexibility, Robustness, Ease of Use, Effectiveness (low overhead)\\ \hline
      
      Kshot \cite{Function7Kshot} & X86 SMM, intel SGX & \checkmark & \checkmark & - & - & Function & Effectiveness (low overhead), Trustworthiness \\ \hline
      
      Safestack \cite{Inst5-chen2013safestack} & Linux & \checkmark & - & \checkmark & \checkmark & Instruction & Efficiency, Safely, Scalability \\ \hline
      Instaguard \cite{Inst4-chen2018instaguard} & Android & \checkmark & - & \checkmark & \checkmark & Instruction &Efficiency \\ \hline
      Replus \cite{Inst6-Function5-Replus-chen2015framework} & Linux & \checkmark & - & \checkmark & \checkmark & Instruction, Function &Efficiency, Lightweight, Practicality  \\ \hline
      DUSC \cite{Function10-Orso2002DUSC} & Java runtime system & - & \checkmark & - & - & Function & Effectiveness (low overhead) \\ \hline
      Vulmet \cite{xu2020automatic-Kernel4F-Inst2Vulmet} &Android & - & - &\checkmark & - & Instruction &
Correctness, Robustness, Efficiency
 \\ \hline
  Seamless \cite{Kernel3Seamless}&  Linux & - &\checkmark & \checkmark & - & Kernel & Effectiveness, Reliability \\ \hline
  Kup \cite{Kernel1Kup} & Linux & - &\checkmark & \checkmark & - & Kernel & Effectiveness, Reliability \\ \hline
      
         Cure \cite{Function8-Cure-zhao2016} & x86  & - &\checkmark &\checkmark &- & Function& Effectiveness, Safety \\ \hline
        Piston \cite{Function10-salls2017piston} & x86 & \checkmark & -  & - & \checkmark & Function& Effectiveness, Persistence \\ \hline
         DynSec  \cite{Inst7-payer2013dynsec} & x86
         & \checkmark & - & \checkmark &\checkmark & Instruction & Correctness,  Effectiveness (low overhead), Efficiency  \\ \hline 
         Gitar \cite{Functiona12ruckebusch2016gitar}  & Contiki, TinyOS  & - & - & - & \checkmark & Function & Efficiency, Sustainability, Maintainability  \\ \hline
       HotAsap\cite{Inst7-payer2013hotAsap}   & Apache web server &
     \checkmark & - & - & - & Instruction & Correctness,  Effectiveness (low overhead), Efficiency \\ \hline
     Karma \cite{Inst1-Karmachen2017adaptive} & Android & \checkmark & - & - &  - & Instruction & Adaptiveness,  Safety, Effectiveness  
     \\ \hline
    Multiverse \cite{Function9-rommel2019multiverse}  & Linux, cPython, musl & \checkmark & - & - & \checkmark & Function & Efficiency, Ease of use, Flexibility, Portability \\ \hline 
    C-Multiverse \cite{function11rothberg2016CompMultiverse}    & Linux, cPython, musl &  \checkmark & - & - & \checkmark & Function & Efficiency, Ease of use, Flexibility, Portability \\ \hline
   BinPatch \cite{Function13hu2019BINPATCH} & Linux, x86 & \checkmark & - & - & -  & Function & Efficiency, Persistence \\ \hline 
   ShawdowPatching \cite{VM1-le2014shadowPatching} & Virtual & - & \checkmark & - & - & VM & Effectiveness \\ \hline
   Proteos \cite{Process3-giuffrida2013safe-Proteos} & x86 Proteos & \checkmark & \checkmark & \checkmark & - & Process & Effectiveness, Scalability, Reliability, Security \\ \hline 
   Waitfree \cite{Process4-rommel2019waitFree} & x86, Linux & - & - & \checkmark & \checkmark & Process & Low overhead\\ \hline 
    Vpatcher \cite{VM2-zhang2014vpatcher} & Virtual & -- & -- & -- & \checkmark & VM & Correctness, Efficiency \\ \hline 
    Orthus \cite{hypervisor2-Orthus-zhang2019fast}  &
      Virtual & -- & \checkmark & - & - & Hypervisor & Efficiency, Effectiveness, Scalability  \\ \hline 
      HyperFresh\cite{hypervisor1-hyperFresh-doddamani2019} & Virtual & \checkmark & \checkmark & - & - & Hypervisor & Effectiveness, Low overhead \\ \hline
    ContainerECU \cite{Container4-ayres2021continuous} & Electronic Control Unit (ECU) & - & - & \checkmark & - & Container & Efficiency, Effectiveness (low overhead), Scalability \\ \hline 
 Cetratus \cite{Process1-mugarza2020cetratus} & x86, PowerPC & - & - & - & \checkmark & Process & Stability \\ \hline
 MVEDSUA \cite{Process5-pina2019mvedsua} & C program & - & \checkmark & - & \checkmark & Function &  Availability, low latency, effectiveness \\ \hline
 Tedsuto \cite{Process6-pina2016tedsuto} & Java program & - & \checkmark & \checkmark & - & Process & Effectiveness, Efficiency\\ \hline
  Kitsune \cite{Process8-hayden2012kitsune, Process8-hayden2014kitsune} & C-program & - &\checkmark & \checkmark & - & Process & Effectiveness (no overhead), Efficiency \\ \hline
 Ginseng \cite{Process9-neamtiu2006Ginseng} & C program & - & \checkmark & \checkmark & - & Process & Scalability,  Effectiveness (no overhead), Efficiency \\ \hline
Polus \cite{Function6-chen2007polus} & x86 & \checkmark &\checkmark & - & \checkmark &  Function & Effectiveness (low overhead), Efficiency \\ \hline
Ksplice \cite{Function1KspliceArnold2009} & Linux kernel & \checkmark & \checkmark & \checkmark & - & Function & Efficiency \\ \hline
Appwrapper \cite{Fucntion2-AppWrapper-lee2020your} & Java program & - & - & - & \checkmark & Function & Efficiency, Effectiveness \\ \hline
OSSPatcher \cite{Function3-OSSPatcher-duan2019automating} & C/C++ program & - & - & - & \checkmark & Function & Efficiency, Feasibility, Variability \\ \hline
HotPatcher \cite{Function4-hotpaccher-jeong2017functional} & ARM, x86\_64 & \checkmark & - & - & \checkmark & Function & Effectiveness (low cost and overhead), Reliability\\ \hline
    \end{tabular}
    \label{tab:study-capabilityl}
    \vspace{-15pt}
\end{table}

\subsubsection{\textbf{Resource Transformation}}

Due to the requirement of continuous operation, a crucial capability of a patching system is to achieve compatibility between currently occurring activities and the expected new behavior. Existing resources (e.g., runtime data objects) may need to be converted to comply with the new code expectations while maintaining compatibility with the existing external users. Note that, particularly in uncooperative patching scenarios, the resources to be transformed must first be determined and located. While locating known objects in RAM can be simplified by utilizing compiler or dynamic linker information, determining the actual list of objects to transform is more complicated.
Furthermore, the resources that need to be transformed to be compatible with the new code depend on the actual change introduced by the patch. In contrast, cooperative solutions may rely on continuous state tracking to aid the new code when necessary. Similar state tracking approaches relying on safe checkpoints are commonly used in intermittent computing 
\cite{lucia2017intermittent}.

Gupta et al. \cite{gupta1996formal} have shown that arbitrary program state transformations are not possible in the general case. Thus, existing solutions are typically limited to particular programming languages and types of transformations only. For instance, only certain types of changes, such as adding a field to a class or changing the number of function parameters, may be supported by some solutions~\cite{Fucntion2-AppWrapper-lee2020your}. Naturally, different technical solutions would be used depending on the level of patch granularity supported and execution environment. 

In cases when most of the compatibility is expected to be retained, simplified resource detachment from old code and reattaching to new code can be enough. For instance, new OS kernels are typically assumed to be compatible with the existing applications. Therefore, as updating the whole OS kernel is not expected to affect running processes, \textit{checkpoint and restore}-type approaches can be employed in practice~\cite{Kernel1Kup, Kernel3Seamless}. These approaches essentially transplant resources from old code to new code. Resource transition from the old code to the new code is referred to as handover if the old code provides some transformation assistance. Conversely, if the new code needs to transform the resource without any old code participation, the operation is referred to as a takeover.

The resources transferred could refer to running OS processes \cite{Kernel1Kup, kernel3Crui} or memory pages of a VM running in a hypervisor \cite{hypervisor1-hyperFresh-doddamani2019}. A significant advantage of resource transplanting is the lack of necessity to modify resources themselves as long as a high degree of compatibility is retained between patches. The most significant downside is the lack of control over external dependencies. For instance, restoring a previously running process to run under a new kernel cannot guarantee that network sockets used by the process would still be open on the remote side. Note that some complicated patches may introduce significant changes that are breaking compatibility with external entities outside of control or area of responsibility of the patch. Notably, introducing incompatible changes to the networking code of a server may render all legacy network clients incompatible, thus causing large-scale severe service disruptions.

\subsubsection{\textbf{Safe Execution State Determination}}
During runtime patch deployment, extra care must be taken to prevent external execution threads from executing or accessing code and data fragments that are being modified. A naïve approach of straightforwardly suspending all other threads (known as stop-the-world) for the duration of the patching process is not adequate as resuming those threads after patching might lead to inconsistencies. Furthermore, suspending running threads may cause noticeable service interruption for more prolonged (i.e., large-scale) patching procedures. Therefore, runtime patching solutions aim to determine time slots that are safe for thread suspension and suspend execution threads for the shortest time possible.

A safe patching time slot refers to a point in software execution where the patched code or data is not in use. For example, if rewriting instructions comprising a function are conducted when the function is being executed, half-changed instructions may be executed, leading to inconsistent behavior. Depending on the execution environment, various safe timing detection algorithms may be implemented in practice.

Two main possibilities are to detect when a certain execution condition is met or rely on existing code assistance when possible. Detection is typically achieved through memory access monitoring and breakpoints mechanism. Specifically, a monitoring stack is commonly used in practice to determine whether the given function is currently being executed \cite{Inst6-Function5-Replus-chen2015framework}. Naturally, relying on existing code assistance is only possible when such assistance facilities are part of the code design and not for legacy software. Automated patch code analysis may reveal critical code points suitable for safe patching and set corresponding breakpoints. Once such breakpoints are triggered, the patching deployment can commence safely. Alternatively, semi-automated patching solutions relying on safe time detection assistance use function hooks to allow notifying the patching system when a safe execution point is reached \cite{Function8-Cure-zhao2016}. In some sense, safe point detection is offloaded to the original software developers and not directly implemented by the patching system.

Interestingly, Intaguard \cite{Inst4-chen2018instaguard} attempts to fix the software behavior only when vulnerability exploitation is detected rather than patching code in advance. Instaguard waits for program execution to reach a vulnerable instruction, then after intercepting the execution, corrects the program behavior and returns execution to the original code. In some sense, rather than looking for \textit{safe} patch time slot only a \textit{necessary} moment is being detected in case of Instaguard \cite{Inst4-chen2018instaguard}. Lastly, some patching solutions side-step the necessity to detect safe time points by following the Co-exist\& Decay strategy. In these situations, patch safety is achieved by isolating different user requests or sessions not to impact each other to avoid potential inconsistencies.

\subsubsection{\textbf{Execution Path Dispatch}}

\begin{figure}[thb]
    \centering
    \includegraphics[scale=0.75]{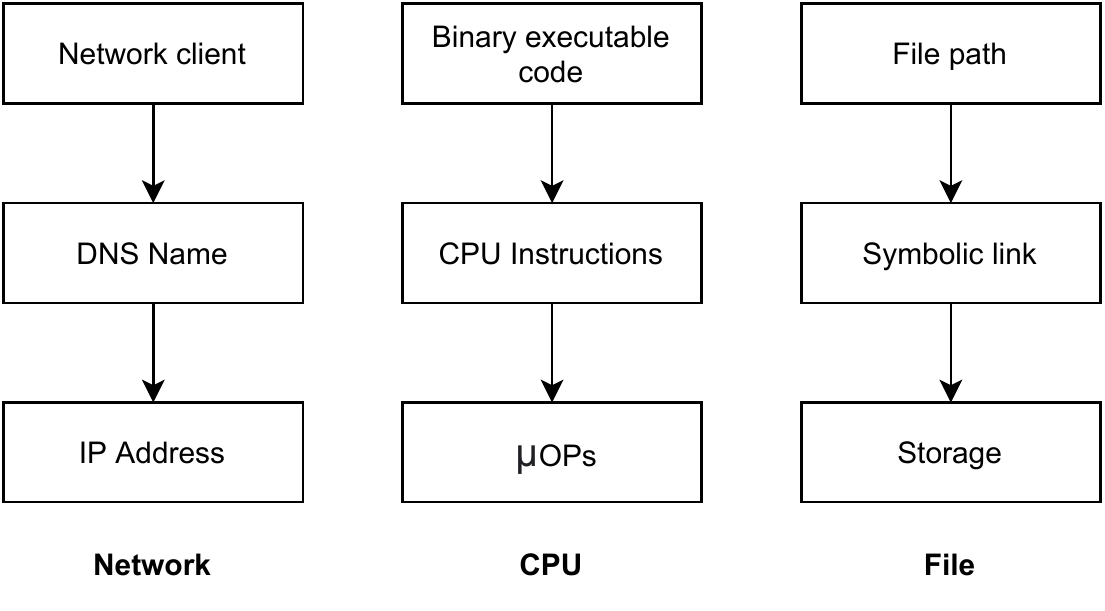}
    \caption{Examples of network, CPU and file-level indirection. $\mu$Ops refers to micro operations of CPU}
    \label{fig:indirection}
    \vspace{-5pt}
\end{figure}

Code execution path dispatching is commonly achieved through various \textit{indirection} layers. This indirection could refer to either code fragments or individual data objects. Generally, \textbf{indirection} is commonly used to achieve loose coupling between different entities. Figure \ref{fig:indirection} illustrates several indirection layers frequently used in modern software systems. Indirection layers such as shown in Figure \ref{fig:indirection} serve multiple purposes, including user convenience. For instance, remembering human-readable DNS names is easier for humans compared to numeric IP addresses. In the context of software patching and updating, the loose coupling achieved through indirection simplifies replacing old entities (e.g., function and instruction) with the patched ones. Specifically, updating the mapping in the indirection layer enables a seamless transition to the new entities. For example, updating DNS records to point to new IP addresses, changing symbolic link destination or redefining CPU instruction (in terms of $\mu$OPs) ensures that future usage would point to the updated entity.

Merely replacing a buggy entity with the fixed one is, however, not enough as old entities might still be in active use by existing execution threads. In the case of long-living usage transactions (i.e., network sessions), the old entities may not be released for prolonged periods. Therefore, various dispatching mechanisms are implemented \textit{in conjunction with indirection layers} to control which of the entities would be used at a given time~\cite{Function9-rommel2019multiverse, function11rothberg2016CompMultiverse}. Such dispatching mechanisms are the core of the co-exist \& decay strategy, with the exact definitions of \textit{entity} and dispatching algorithms used widely varying depending on the patching granularity supported.

Explicit version numbers are commonly used for
dispatching\footnote{\url{https://docs.microsoft.com/en-us/aspnet/core/grpc/versioning?view=aspnetcore-6.0}}$^,$\footnote{ \url{https://github.com/grpc/grpc/blob/master/doc/versioning.md}}. Such explicit versioning ensures compatibility between different copies of the co-existing code or data and external users. Due to the associated performance and memory overheads, multi-version dispatching may be used as a temporary measure while a developer is working on a longer-term solution~\cite{Function9-rommel2019multiverse, function11rothberg2016CompMultiverse}. In addition, to minimize at least the memory overhead, the gradual decay (i.e.,~cleanup) of old code and resource may be implemented as a feature of some patching solutions.

\section{Runtime Patching Responsible Entities} \label{sec:PatchResponsibility}

\begin{figure}[thb]
     \centering
     \begin{subfigure}[b]{0.3\textwidth}
         \includegraphics[width=\textwidth]{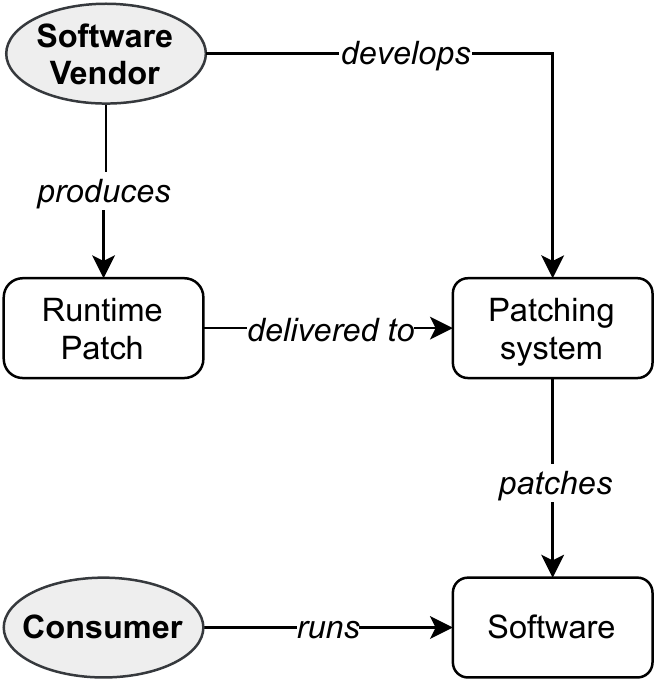}
         \caption{Vendor assisted}
         \label{fig:VendorAssisted}
     \end{subfigure}
     \hfill
     \begin{subfigure}[b]{0.3\textwidth}
         \centering
          \centering
         \includegraphics[width=\textwidth]{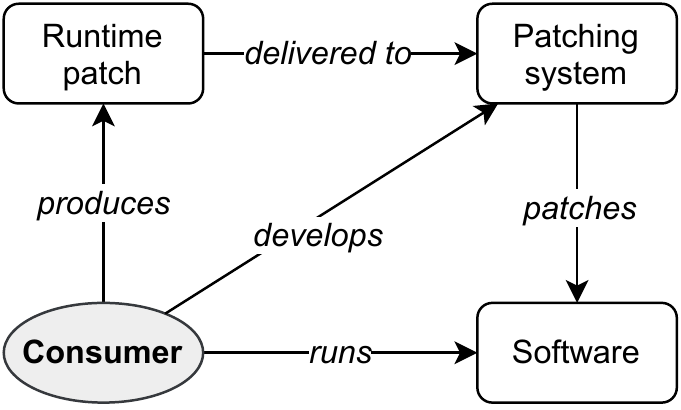}
         \caption{Consumer assisted}
         \label{fig:ConsumerAssisted}
     \end{subfigure}
     \hfill
     \begin{subfigure}[b]{0.3\textwidth}
         \centering
          \centering
         \includegraphics[width=\textwidth]{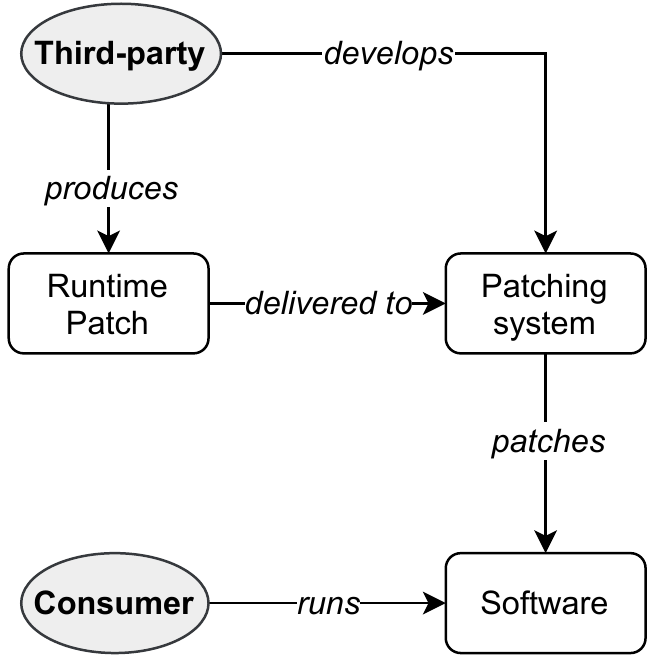}
         \caption{Third party assisted}
         \label{fig:ThirdPartyAssisted}
         \end{subfigure}
        \caption{System overview of runtime patching system with different responsible entities  - (a) software vendors, (b) consumers and (c) third parties. Third parties refer to patch system developers who are not vendor or consumer. The oval shapes in the diagram represent human entities and the square shapes represent software entities. }
        \label{Fig:PatchResponsibility}
\end{figure}

The entities and their roles involved in the runtime patching process are another important aspect of the patching workflow that is not commonly explicitly discussed. We identify three key parties involved in the typical patch life cycle, which are (i)~the original software vendor, (ii)~software end-user/consumer and (iii)~third party patch system developer. Figure \ref{Fig:PatchResponsibility} depicts an overview of three different scenarios of patch deployment with respect to different parties and their corresponding responsibilities. Consider that the same entity could fulfill all three roles in simple cases. For instance, for an in-house developed software used within company boundaries only, the company develops software (along with associated patches), uses the software and provides further maintenance through implementing patches as required. Note that we explicitly focus on patch deployment rather than patch development or automated generation responsibilities. However, the framework Replus has considered both developer and consumer perspectives of patching and divided the patching duties between developers and consumers~\cite {Inst6-Function5-Replus-chen2015framework}. Table~\ref{tab:PatchResponsibility} provides a mapping of various granularity with patch responsibility proposed in the literature review. 

\begin{table}[htb]
    \centering 
     \caption{Analysis of Existing Patching Solution with Patch Responsibility}
    \begin{tabular}{p{1.8cm}|p{4cm}|p{3cm}|p{5.5cm}}
    \midrule
    \textbf{ Granularity}  & 
     \textbf{ Vendor} & 
      \textbf{Consumer}& \textbf{Third party} \\ \midrule
      Instruction   & - & DSU \cite{inst8-DSU-Hics2005}& DSU~\cite{inst8-DSU-Hics2005}, Safestack~\cite{Inst5-chen2013safestack}, InstaGuard~\cite{Inst4-chen2018instaguard}, Replus~\cite{Inst6-Function5-Replus-chen2015framework}, Vulmet~\cite{xu2020automatic-Kernel4F-Inst2Vulmet}, Wordpatch~\cite{Inst3-WordPatch-chamith2016living} \\ \hline
      Function & Multiverse~\cite{Function9-rommel2019multiverse}, CompMultiver~\cite{function11rothberg2016CompMultiverse}, Gitar~\cite{Functiona12ruckebusch2016gitar}  &  Ksplice~\cite{Function1KspliceArnold2009}, Piston~\cite{Function10-salls2017piston}, UpdateCalculus~\cite{Function14-bierman2003UpdateCalculus} & Replus~\cite{Inst6-Function5-Replus-chen2015framework}, Polus~\cite{Function6-chen2007polus}, Cure~\cite{Function8-Cure-zhao2016}, Appwrapper~\cite{Fucntion2-AppWrapper-lee2020your}, OSSPatcher~\cite{Function3-OSSPatcher-duan2019automating}, HotPatcher~\cite{Function4-hotpaccher-jeong2017functional} \\ \hline
      Process  &   Katana~\cite{Process7-ramaswamy2010katana}, Proteos~\cite{Process4-rommel2019waitFree},  MVEDSUA~\cite{Process5-pina2019mvedsua},  Ginseng~\cite{Process9-neamtiu2006Ginseng} &  Cetratus~\cite{Process1-mugarza2020cetratus}, Proteos~\cite{Process3-giuffrida2013safe-Proteos} & Tedsuto~\cite{pina2016tedsuto}, Kitsune~\cite{Process8-hayden2012kitsune, Process8-hayden2014kitsune}\\ \hline
      VM  & - & -&  ShadowPatching~\cite{VM1-le2014shadowPatching}, Vpatcher~\cite{VM2-zhang2014vpatcher}\\ \hline
      Hypervisor & Orthus~\cite{hypervisor2-Orthus-zhang2019fast}, HyperFresh~\cite{hypervisor1-hyperFresh-doddamani2019} & - & -\\ \hline
      Container & - & - & Opatch~\cite{Container1-Opatch-tunde2020}, SelfPatch~\cite{Container2-tunde2020selfPatch}\\ \hline
      Kernel  & - & - & Seamless~\cite{Kernel3Seamless}, Kup~\cite{Kernel1Kup} \\ \hline
     
    \end{tabular} 
    \label{tab:PatchResponsibility}
    \vspace{-5pt}
\end{table}

\subsection{Vendor-assisted Patching} 

In vendor-assisted patching scenarios, the software is designed with runtime patching support in mind, with end-users having little to no involvement in the patching process~\cite{Function9-rommel2019multiverse, function11rothberg2016CompMultiverse, ruckebusch2016gitar, Process1-mugarza2020cetratus}. With modern software commonly enabling automated patching by default, users at most have an option to apply or skip a given patch. Original software vendors have deep knowledge of how the software operates, which (in principle) enables developing complex and safe patches. Also, having access to the software sources opens up several avenues not available for end-users or third parties in closed-source software. For instance, runtime patches could be automatically generated based on the source code changes conducted by the software developers. Having control over software development, deployment and further management enables designing tightly integrated patching solutions~\cite{ Process3-giuffrida2013safe-Proteos}. Patches can be developed, delivered to end-users and automatically deployed by the vendor as shown in Figure~\ref{fig:VendorAssisted}, which significantly reduces the burden of management on the end-users. 

Two potential downsides of vendor-assisted patching are low applicability for legacy systems and lack of flexibility from an end-user perspective. Legacy systems lacking a run-time patching subsystem would fall victim to potential vulnerabilities. Completely redesigning such legacy systems might be prohibitively expensive or impractical. End users would also be entirely dependent on the willingness of the software vendors to fix a given issue. Lower priority or site-specific problems may not receive enough attention from the developers, causing a lack of patches addressing some of the end-user needs.

\subsection{Consumer-assisted Patching}
 
In some cases, software consumers may opt to manage patch deployment for the software they use, which we refer to as consumer-assisted patching. Three possible reasons include a higher level of control, more flexibility and lack of support from the original software vendor. At the expense of increased labor required, implementing own patching strategy essentially allows end-users to gain a higher level of control over the software system as shown in Figure~\ref{fig:ConsumerAssisted}. Consequently, fine-grained or user-specific patches can be deployed by the consumers based on their current and ever-changing requirements~\cite{Function10-salls2017piston, Process8-hayden2012kitsune, Process8-hayden2014kitsune}. Also, implementing own patching solutions may be the only option for legacy systems that do not include runtime patching support or are no longer maintained.

However, achieving the desired flexibility may not be easy for the end-users as they lack intimate knowledge of data structures and flow used within the software. Due to the complexities in foreseeing the potential side-effects of the patch-induced changes, applying potentially inconsistent/faulty patches poses significant risks. The complexity of predicting potential patch failures or discrepancies is higher for closed-source software while being somewhat more manageable for fully open-source software. In a subset of situations, end-users might limit their efforts to converting traditional source- or binary-level patches into a dynamic form rather than developing their patches from scratch. In other words, end-users might focus on developing a runtime patching system to deploy the traditional vendor-produced offline patches in a dynamic manner.

\subsection{Third-party-assisted Patching}

Third-party runtime patching systems refer to solutions that are applicable in a wider set of software environments and represent a practical trade-off between vendor and consumer efforts. Not possessing the knowledge of inner software details and lacking information on individual end-user needs, third parties aim to achieve some level of generality in terms of applicability in different contexts such as specific programming languages, OS or certain patch granularity. For instance, some solutions support solely Java- or C-based applications~\cite{Function10-Orso2002DUSC, Inst7-payer2013hotAsap, Process8-hayden2014kitsune}, while others focus on specific OS environments~\cite{Process3-giuffrida2013safe-Proteos,Inst4-chen2018instaguard}.

The third-party-assisted patching approaches do not expect assistance from the old code while still simplifying patch deployment. One common theme within third-party efforts is to focus on converting existing source-based vendor-developed offline patches to be applied at run time by end-users. For this purpose, third-party-assisted software updating implements an external entity that would replace old code with the new code while preventing potential service interruptions. For instance, the proposed approach by Chen et al.~\cite{Inst5-chen2013safestack} requires the installation of a runtime patch applicator in the online production system. Figure~\ref{fig:ThirdPartyAssisted} shows a high-level overview of third-party-assisted patching. Most of the research efforts we have studied are geared towards third-party assisted patching~\cite{ xu2020automatic-Kernel4F-Inst2Vulmet, Inst4-chen2018instaguard, Fucntion2-AppWrapper-lee2020your, Function3-OSSPatcher-duan2019automating, Function6-chen2007polus, Function8-Cure-zhao2016, Process6-pina2016tedsuto, hypervisor2-Orthus-zhang2019fast, Container2-tunde2020selfPatch, Kernel1Kup}.

\section{Discussion}

The high performance and memory overheads primarily define the runtime patching viability as opposed to traditional offline patching. These overheads are caused by the sheer size of the execution state inherent to modern complex software and difficulties in determining safe update points. In mission-critical domains such as medical, military or rescue operations, it is essential to maintain uninterrupted service operations. Highly critical systems with plentiful resources can tolerate high performance or memory overheads common for runtime patching systems. However, the resource-constrained systems such as embedded IoT devices may require more efficient patching solutions with lower performance or memory overheads. Our analysis of the reviewed studies hints that vendor-provided patches have more potential to reduce the associated overheads. The intimate knowledge of the software internals along with the collaborative nature of software, simplify many of the patching steps. Collaborative nature can refer to requesting controlled code execution suspension instead of abrupt code interruption or lengthy waiting of quiescent state. Similarly, the existing old code can collaborate in providing the list of resources in use rather than requiring complicated manual resource tracking.

According to our proposed taxonomy, we also conclude that runtime patching is dominant in open-source ecosystems, with the majority of patching frameworks designed to be used by independent third parties as opposed to software vendors and consumers directly (as shown in Table~\ref{tab:PatchResponsibility}). This is likely explained by software vendors not paying enough attention to runtime patching functionality. An intermediate middle-man role seems to be forming, focused on service continuity. Cloud infrastructures are a typical example where service continuity is not a separate solution exposed to end users by software vendors but is instead integrated into the cloud offering by the cloud provider. Access to source code significantly simplifies the understanding of application internals, enabling the development of efficient runtime patching solutions suitable for a given application. Not limiting a patching solution to particular vendors or end-users also contributes to a patching system’s popularity due to its broader applicability.

We observe a common commercialization trend in runtime patching solutions where rather than monetizing the patching system itself, actual patches are provided as part of paid subscription. For example, the existing commercial solutions such as KernelCare by TuxCare\footnote{\url{https://tuxcare.com/live-patching-services/}}, Ksplice by Oracle\footnote{\url{https://ksplice.oracle.com/}} and kpatch by Red Hat\footnote{\url{https://access.redhat.com/solutions/2206511}} provide patches that are possible to apply at run time. Customers essentially pay for adapting and supporting regular source-level changes to runtime patch format. Note that some types of changes, such as semantic structure modifications or specific system functions replacement, may be explicitly unsupported by certain patching solutions\footnote{\url{https://github.com/dynup/kpatch}}\textsuperscript{,} \footnote{\url{https://manpages.ubuntu.com/manpages/trusty/man8/ksplice-create.8.html}}.

We further observe that despite these successful commercial solutions focusing on Linux kernel patching, the actual granularity used internally is not necessarily kernel-level but can be implemented at instruction or function levels (Tables \ref{tab:instPatching}-\ref{tab:kernelPatching}). Notably, we categorize Ksplice into function-level granularity. Consequently, depending on the patch design choices, different capabilities, as shown in Table \ref{tab:study-capabilityl}, may need to be implemented. For instance, Ksplice opted not to implement execution code path dispatching capability while supporting the other three patch implementation capabilities identified by our taxonomy.

Closed-source commercial runtime patching solutions are predominantly developed, controlled and managed by the corresponding software vendors. This is because only vendors have access to software source code, which significantly simplifies patch development and preparation. Consequently, such vendor-curated patching solutions are tightly integrated into software and not provided separately for a wider audience. For instance, Microsoft employs a run-time patching solution to minimize Azure SQL Database engine service disruptions during patching\footnote{\url{https://techcommunity.microsoft.com/t5/azure-sql-blog/hot-patching-sql-server-engine-in-azure-sql-database/ba-p/849700}}. According to their statistics, over 80\% of SQL bugs can be remediated using this patching system. As per our taxonomy, this approach is a function-level coexist and decay solution. Despite being closely tied to a particular software application, a similar compiler-assisted approach can be used for other Visual C-based applications due to the ubiquity of patch loading and function redirection steps. Another commercial solution we found, namely MuleSoft CloudHub platform\footnote{\url{https://docs.mulesoft.com/cloudhub-2/ch2-patch-updates}}, primarily focuses on maintaining continuous cloud service. Due to the closed nature, the provided patching solution description is scarce on implementation details; however, according to our taxonomy falls under process-level coexist and decay solutions.

Note that while some dynamic programming languages such as Python or JavaScript provide various dynamic code loading mechanisms, they do not directly map to actual patch contents. A developer must still prepare the code that will conduct necessary state transformations by writing a piece of code that will be loaded and applied. Nevertheless, further development of dynamic code handling is a promising path toward efficient runtime patching system implementation.

\section{Open Challenges and Future Research Directions in Runtime Patching}  \label{sec:futuredirection}

A number of observations related to gaps in the existing knowledge and efforts can be made based on the conducted review. In short, we notice that despite significant research and development efforts, runtime patching is still underutilized in practice. For example, an amusingly recent Log4J vulnerability was proposed to be patched by exploiting the vulnerability itself\footnote{\url{https://www.lunasec.io/docs/blog/log4shell-live-patch/}}. Furthermore, despite individual software applications using various runtime patching solutions, a lack of general solutions indicates a significant research potential in this domain, specifically developing compiler- and OS kernel-augmented approaches.

\begin{figure}[htb]
    \centering
    \includegraphics[scale=0.8]{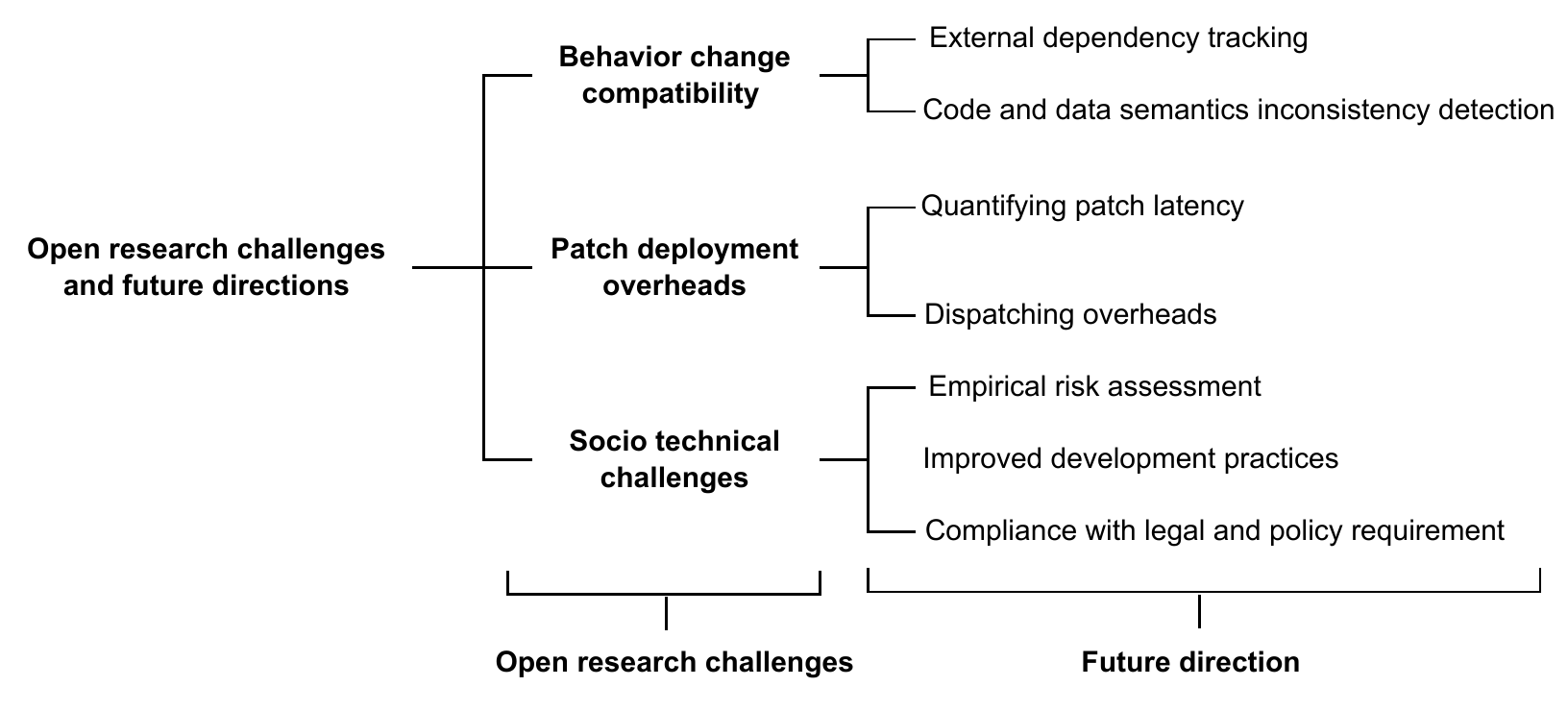}
    \caption{Open research challenges and potential future directions for runtime patching }
    \label{fig:OpenIssues}
    \vspace{-5pt}
\end{figure}

We have identified a number of open issues and potential future directions that are shown in Figure~\ref{fig:OpenIssues} and discussed in the following subsections. Firstly, most of the existing solutions are narrow-scoped and severely limited in their applicability. For instance, a runtime patching solution may employ specific language or execution platform or OS features, restricting the usefulness of such a solution. Alternatively, only specific types of changes (e.g., adding a class method) may be supported by a patching solution, posing challenges for non-trivial modifications.

Secondly, the lack of a systematic approach in runtime patching could be explained by the ad-hoc nature of patch development currently occurring in practice. Unlike traditional software development, which is well established with a multitude of development practices already matured, patch development looks less organized. This is because patches are typically developed under tight time pressure due to the associated security risks. High-profile vulnerabilities can cause high monetary losses, incentivizing the quickest possible patch deployment. This leaves little to no time to approach the task systematically, leading to situations when patches are shortly followed by patches-for-patches (sometimes more than once) to fix the original patches that turned out to be buggy\footnote{\url{https://arstechnica.com/information-technology/2021/12/patch-fixing-critical-log4j-0-day-has-its-own-vulnerability-thats-under-exploit/}}$^,$\footnote{\url{https://blog.malwarebytes.com/exploits-and-vulnerabilities/2021/08/microsofts-printnightmare-continues-shrugs-off-patch-tuesday-fixes/}}.

Lastly, runtime patching capability often comes as an afterthought in contrast to traditional source-level patching. Runtime patches may not even be developed independently but may be automatically generated from developer-supplied source code changes. This leads to a shortage of cooperative patching solutions that could simplify further patching through various degrees of assistance from the existing codebase. Developing a cooperative updating subsystem requires additional efforts at design and development phases. It is thus not commonly being used in practice. Furthermore, foreseeing all possible patching and resource handover scenarios may not be feasible, rendering such implementations insufficiently flexible in the long term. 

\subsection{Behaviour Change Compatibility}

Compatibility between in-flight activities and updated code needs to be maintained to ensure user experience continuity. Patches that introduce significant externally observable changes may lead to incompatibilities with the expectation of external observers. Therefore one of the tasks of a run-time patching system is to achieve code and data compatibility or, at the very least, detect potential incompatibilities prior to patch deployment. Dealing with potential incompatibilities between original and patched code requires determining the \textit{scope} (i.e., how far are the patch effects observable) and \textit{type} (i.e., what becomes incompatible) of the incompatibility. For instance, the patch application might be stopped entirely or postponed if certain incompatibilities are detected. Detecting the most disruptive patches can aid in reducing the associated downtime and service interruptions. Considering the mentioned issues of behavior change compatibility, we further discuss the future research directions potentially valuable for solving different aspects of such issues between original and patched code. 

\subsubsection{\textbf{External Dependencies Tracking}} 

Determining where a potential incompatibility lies requires tracking the code's various external dependencies that also need to be updated. For instance, if a function-level patch alters the arguments accepted by the function, all external code fragments that call the function need to be updated accordingly. Moreover, recursive analysis of further code fragments depending on the code that calls patched function is required to propagate the changes as necessary. Similarly, patching a network service application might require updating remote client-side software to retain network-level service compatibility. This process is similar to recursive taint tracking commonly used to track the flow of untrusted user input to identify and rectify unsafe input uses \cite{ermolinskiy2010towards}. Generally speaking, security-oriented patches are attempted to be minimized not to change the code behavior drastically. However, it can be argued that with tightly integrated software, even a tiny patch (function- or instruction-level) can lead to drastic side effects. Therefore, understanding where the patch impact would be visible is a crucial preliminary step in runtime patching. Hence, researching and developing quantitative \textbf{patch impact estimation} techniques would be helpful in order to detect potential external incompatibilities. 

Further patch impact estimation techniques may be used to estimate a given patch's potential disruptiveness. Such complexity metric would enable formally verifying the impact of a given patch. Therefore, simple patches that do not affect end-users can be applied immediately, whereas applying complex and disruptive patches would require planning a downtime. Precisely, the externally-observable software behavior changes would be estimated to determine a given patch impact. Given the increasing trends in software isolation (through containerization and virtualization), special attention needs to be paid to network-related patch side-effects that are observable even on remote network nodes.

\subsubsection{\textbf{Code and Data Semantics Inconsistency Detection}} 

When a developer creates the new code fragment, some inconsistencies may be introduced inadvertently. Simple inconsistencies, such as data type mismatch can be detected automatically through type safety preserving approaches. Existing solutions typically attempt to achieve type safety through tracking variable data type changes. In contrast, semantic differences (such as the changed meaning of the variable of the same type) may not be possible to be discovered solely by looking at the changed code fragment. For example, changing an integer variable to a string variable can be detected and rectified accordingly~\cite{Function14-bierman2003UpdateCalculus}. 

However, consider a patch that does not alter the variable type (i.e., an integer still stays an integer) but interprets the variable value differently (such as treating meters as feet). Generally, measurement unit handling is prone to such mistakes, and the same distance/angle/weight variable could be interpreted in different ways (e.g., meters vs. feet, degrees vs. radians, pounds vs. kilograms, etc.). Furthermore, a patch that performs a transition from one measurement system to another may not cause a type mismatch error (i.e., violate the type safety) but could still cause issues further down the execution chain. Detecting such semantic changes is unquestionably more complicated as all of the related code paths that rely on the variable outside the scope of the patch must also be analyzed. Unfortunately, no feasible solution aiming to detect semantic inconsistencies has been proposed so far. Perhaps meta-languages similar to $\mu$DSU~\cite{Fucntion14cazzola2018muDSU} can be adapted to allow outlining a semantic application layer.



\subsection{\textbf{Patch Deployment Overheads}}

In addition to the technical implementation issues, runtime patching overheads must be addressed. These overheads are mainly concentrated in resource (e.g., RAM) and time domains. Coexisting-based solutions imply that multiple copies of code or data must simultaneously be present in memory. Such coexistence might not be an issue for smaller patches; however, applying multiple patches over longer periods may eventually pose challenges. Similarly, resource-constrained IoT devices may also render some solutions impractical due to a lack of hardware resources. For instance, some microchips based on Harvard architecture may not allow direct code modification, leading to the necessity to re-flash main memory contents and restart the execution. Note that, even in such restricted devices, certain extensions have been implemented to aid self-modification that would be useful for patching purposes\footnote{\url{http://ww1.microchip.com/downloads/en/AppNotes/doc1644.pdf}}. Architecture-, device- and resource-constrained overheads need to be predicted to determine patch suitability for a given system real-time requirement. We identify and discuss the associated quantitative metric research directions below.

\subsubsection{\textbf{Quantifying Patch Latency}}

\begin{figure}[htb]
    \centering
    \includegraphics[scale=0.6]{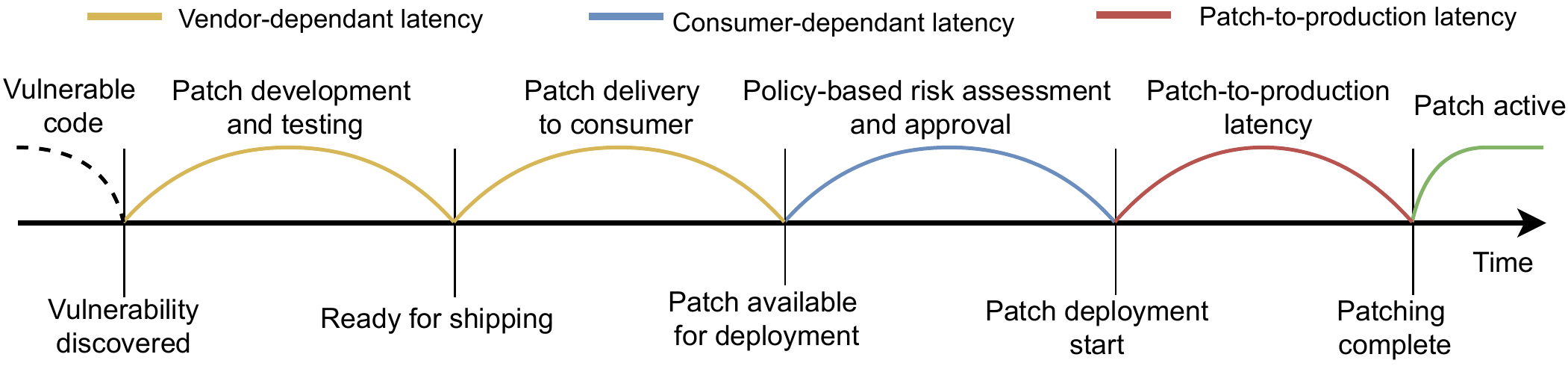}
    \caption{Patch Latency Timeline}
    \label{fig:timeline}
\end{figure}

Different patch latency metrics can be defined because of many factors involved in patch preparation, deployment and testing. Figure~\ref{fig:timeline} shows the different types of patch latency that might be encountered in practice. The \textit{longest time-frame} typically used as a metric of patch refers to the time passed between the discovery of vulnerability or bug and actual update in production. In severe cases, \textit{vendor-to-production latency} can be in the order of magnitude of days to months\footnote{\url{https://www.businessinsider.com.au/zoom-security-flaw-hackathon-dropbox-2020-4?r=US&IR=T}}. This latency, however, does not strictly depend on technical issues and could be dominated by other organizational-level inefficiencies such as supply chain delays, internal company policies, vendor policies and the priority of the system.

Setting aside policy- and development-induced delays, the \textit{patch-to-production }latency metric can be used while considering the technical aspects of runtime patching. This latency essentially covers patch deployment and patch activation steps. However, determining patch-to-production latency numerically is not straightforward, as different strategies cause delays in different aspects. For instance, some solutions may opt to implement co-exist \& decay strategy in order to avoid any noticeable system downtime. This, however, means that currently running interactions may not notice the effects of the patch immediately (\textbf{activation latency}). Alternatively, state transformation strategies ensure that effects of the patch are observed immediately at the expense of briefly interrupting running interactions (\textbf{critical transformation latency}). A more systematic approach in quantifying runtime patching latencies would be beneficial for evaluation and direct comparison of different patching implementation approaches.

\subsubsection{\textbf{Dispatching Overheads}}

Code path or resource dispatching approaches attempt to minimize software system overhead. However, even lightweight dispatching overheads tend accumulate over multiple repeated patches causing increased latencies. While the \textit{decay} phase is designed to release an unused resource, both additional memory and time requirements may grow for long-living services. In other words, if external entities keep using the resource for a prolonged period of time, the resource would not be released, leading to a higher amount of resources such as memory, network sockets or file descriptors being used. While focusing on immediate gains, some existing solutions tend to overlook some long-living software usage scenarios during their evaluation to estimate the effects of accumulated overheads~\cite{Function6-chen2007polus}. Investigating dispatching overheads in relation to software system usage under high load would provide a better understanding of patching implementation applicability in practice.

\subsection{Socio-technical Challenges}

In addition to purely technical issues related to runtime patching, several socio-technical aspects may pose additional complications. For instance, delays due to imposing internal policies or legislation may significantly slow down the patching procedures. Organizations may consider improving software development practices that incorporate runtime patching at the design phase. However, the monetary and time costs associated with rewriting large legacy software packages according to improved development practices may drive off software developers. Furthermore, in some cases, extra complications in patching processes may be induced by supply chain participants (such as mobile service providers or hardware vendors) for the purposes of patch verification or secure distribution. We identify some potentially beneficial research directions to solve socio-technical challenges hindering wider adoption of runtime patching.

\subsubsection{\textbf{Empirical Risk Assessment}}

Empirical evaluation of the potential risks of deploying runtime patches requires significant attention from a business perspective. Organizations need to predict potential patch consequences regarding service disruption, data incompatibility, impact on adjacent services or even new vulnerabilities introduced by a patch. While ultimately, business analysts may make the final decision manually, some automated quantitative metric determination algorithms can simplify their work. For instance, automatically deducing the list of third-party software services that might be affected by a patch or predicting the service disruption length would greatly aid in making an informed decision backed by actual data. Therefore, researching and developing quantitative patch impact prediction and estimation algorithms at a business scale would be helpful in practice. In addition, another potentially beneficial research direction is the optimization of concrete risk assessment procedures employed in a given company to minimize lengthy delays caused due to following potentially complicated internal policies~\cite{Nesara2022106771}.

\subsubsection{\textbf{Improved Development Practices}}

A group of socio-technical challenges related to runtime patching arises from economic factors such as monetary and time constraints. Notably, some existing solutions implement improved development practices, programming languages or supporting frameworks that significantly simplify runtime patching at the technical level. However, these solutions did not attract highly due to the hidden (non-technical) costs. Firstly, using these solutions requires learning new tools. Secondly, achieving compatibility between existing legacy code and the patching framework requirements typically requires a certain level of code modification. For instance, identifying and defining safe update points within existing code for a developer may be complicated in high-load applications. Combining these two reasons typically lead to high costs of employing such solutions in practice. Thus, a comprehensive study aimed at software development practitioners in different fields might provide additional clues to increasing the adoption of design-time runtime patching solutions. Another pragmatic aspect to explore is the potential avenues in popularising such design-time patching approaches among a wider audience, specifically targeting software developers.

\subsubsection{\textbf{Compliance with Legal and Policy Requirements}}

Lastly, the least technical group of issues complicating runtime patching revolve around external factors like local legislation and external business partners' interaction. Patch delivery and distribution channels may impose organizational complications and significant delays. For instance, patching Android OS needs to overcome organizational vendor-imposed limitations such as obtaining digital signatures from all parties involved in the supply chain~\cite{Inst4-chen2018instaguard}. In addition, highly regulated organizations such as health-, finance- or defense-related may have additional restrictions imposed by specific legislation and regulations\footnote{https://techcrunch.com/2022/01/05/ftc-legal-action-log4j/}, requiring to conduct patching strictly according to predefined procedures. While being out of the scope of this study, legislation and regulation improvement may also benefit from the body of knowledge generated by the patch impact prediction research line.

\section{Conclusion} \label{sec:conclude}

Runtime patching techniques and approaches require radical transformation following the emerging technologies in safety and mission-critical system adoption. Deploying and activating patches with minimum overhead and downtime are crucial steps of applying patches at runtime. We present a taxonomy highlighting the four key aspects that need consideration for runtime patch deployment. We identify and analyze the seven patch granularity levels along with two general patching strategies (state transformation and co-exists \& decay) and three responsible entities (vendor, consumer and third parties). We further define a high-level workflow (Figure~\ref{fig:runtimeWorkflow}) of applying and activating patches supporting the four identified key implementation capabilities that are (i) memory management,(ii) resource transformation, (iii) safe execution state determination and (iv) execution path dispatching. Finally, the review highlights the challenges and suggests potential future prospects. We envision the area mainly needs a systematic approach to develop practical and convenient patching solutions applicable in a wide variety of infrastructures, programming languages and execution environments. The area further requires defining quantifiable metrics to identify the impact of a patch by semantics and dependency tracking solutions. We observe many open opportunities and concerns that need to be considered by vendors, end-users and third parties.

Future research lines include obtaining patching-related real-world data through a practical field study. Analyzing this data would enable understanding the statistics of existing patching challenges in various critical domains. Notable research questions should focus on the patching procedures commonly applied in practice, patch granularity distribution, the complexity of the patches applied, patch-induced system downtime and service continuity, as well as potential incompatibility issues. To conclude, an in-depth understanding of the issues and concerns of practitioner and business requirements is crucial to bridge the gap between runtime patching research and practical implementation/deployment in production.

\section*{Acknowledgment}

The work has been partially supported by the Cyber Security Research Centre Limited whose activities are partially funded by the Australian Government’s Cooperative Research Centres Programme.


\bibliographystyle{IEEEtran}


\bibliography{Main.bib}

\end{document}